\documentclass[prd,preprint,nofootinbib,showpacs]{revtex4}

\usepackage{amsmath}
\usepackage{graphicx}
\usepackage{amsfonts}
\usepackage{latexsym}
\usepackage{amssymb}
\usepackage{epsfig}    
\usepackage{color}

\newcommand{\dis}[1]{\begin{equation}\begin{split}#1\end{split}\end{equation}}
\newcommand{\beqa}[1]{\begin{eqnarray}#1\end{eqnarray}}
\newcommand{\be}{\begin{equation}}
\newcommand{\ee}{\end{equation}}
\newcommand{\eq}[1]{Eq.~(\ref{#1})}
\newcommand{\bfrac}[2]{{\left(\frac{#1}{#2} \right)  }}
\newcommand{\VEV}[1]{{\langle{#1}\rangle  }}
\newcommand{\treh}{T_R}
\def\sstar{\sigma_{*}}
\def\os{{\textrm os}}
\def\sos{\sigma_{\textrm os}}
\def\stwo{\sigma_{\textrm 2os}}
\def\sv{\sigma_{v}}
\def\RD{{R}}
\def\tr{{\tilde{r}}}

\newcommand{\fnl}{f_{\rm NL}}
\newcommand{\gnl}{g_{\rm NL}}

\newcommand{\Mp}{M_P}

\newcommand{\gev}{\,\textrm{GeV}}

\begin{document}

\title{
 Non-Gaussianity and gravitational wave background
 in curvaton with a double well potential
}

\author{Ki-Young Choi}
 \affiliation{
 Department of Physics,
 Pusan National University, 
 Busan 609-735, Korea
}

\author{Osamu Seto}
 \affiliation{
 Department of Architecture and Building Engineering,
 Hokkai-Gakuen University,
 Sapporo 062-8605, Japan
}

%

\begin{abstract}
%
We study the density perturbation by a curvaton 
 with a double well potential and estimate 
 the nonlinear parameters for non-Gaussianity and
 the amplitude of gravitational wave background generated during inflation.
The predicted nonlinear parameters strongly 
 depend on the size of a curvaton self-coupling constant 
 as well as the reheating temperature after inflation 
 for a given initial amplitude of the curvaton.
The difference from usual massive self-interacting curvaton
 is also emphasized.
%
\end{abstract}

\pacs{95.85.Bh, 98.80.Es, 98.80.Cq}

\preprint{PNUTP-10-A09, HGU-CAP 004} 

\vspace*{3cm}
\maketitle


\section{Introduction}

Cosmic inflation solves various problems
 in the standard Big Bang cosmology such as flatness, the horizon, and the
 monopole problems~\cite{Inflation}.
Simultaneously,
 the quantum fluctuation of a light scalar field, e.g., inflaton field $\phi$, 
 generated during inflation 
 is stretched by the rapid cosmic expansion 
 and provides the seed of large scale structure 
 in our Universe~\cite{InflationFluctuation}.

The density perturbation generated in a single field inflation model
 is scale-invariant and almost Gaussian with
 the corresponding nonlinearity parameter $f_{\rm NL}$ 
 much less than unity~\cite{Maldacena:2002vr}.
This is consistent with the current limit on the local type non-linearity 
 parameter $\fnl$ from the 
 Wilkinson Microwave Anisotropy Probe (WMAP) seven-year data,
 $-10<\fnl<74$ at the 95\% confidence level~\cite{Komatsu:2010fb}. 
It is expected that the observational sensitivity is gong to improve significantly 
 by the Planck data~\cite{Planck} and using large scale structure data within the near future.
The non-Gaussianity could be an important observable
 to discriminate between various mechanisms of density perturbation generation.

On the other hand, beyond the simple canonical single field slow-roll inflation,
 the large non-Gaussianity with different shape are 
 generally predicted~\cite{Chen:2006xjb,DBI}. 
There are many models
 for the generation of the observed density 
 perturbation and a large non-Gaussianity.
This can happen during
 inflation~\cite{Zaldarriaga:2003my,Choi:2007su,Byrnes:2008wi,Byrnes:2008zy}, 
 at the end of inflation~\cite{Lyth:2005qk,Alabidi:2006wa,Sasaki:2008uc,Naruko:2008sq},
 preheating~\cite{preheating}, or deep in the radiation dominated era~\cite{Byrnes:2010em}.

The last case includes the ``curvaton''
 scenario~\cite{Mollerach:1989hu,Linde:1996gt,CurvatonLW,CurvatonMT,CurvatonES},
 where the scalar field is too light to make effects around the inflationary epoch
 but it might play an important role much later in the early Universe.
As the Universe expands, the cosmic expansion rate $H$ becomes comparable to the mass
 of the curvaton and the curvaton field starts to oscillate in the radiation
 dominated era. 
After that when the expansion becomes less than the decay rate of the curvaton,
 it decays to light fields and 
 the isocurvature perturbation of the curvaton field becomes adiabatic
 or mixed with that from the inflaton field.
If the curvaton energy density is subdominant at its decay time,
 the large non-Gaussianity is generated in general~\cite{CurvatonNG}.

Another important measure 
 is the gravitational wave background produced during inflation~\cite{GWbackground}
 parametrized by the tensor-to-scalar ratio $r_T$, 
 because it could directly indicate the energy scale of inflation.
$r_T $ is related with one of the inflaton's slow roll parameter $\epsilon$ as 
 $r_T=16\epsilon$, 
 while the density perturbation ${\cal P}_{\zeta} \propto \epsilon^{-1}$
 in a single field inflation model. 
On the other hand, in the curvaton scenario,
 the density (scalar) perturbation comes from the curvaton.
Nevertheless the (non-)observation of $r_T$ gives strong constraint on the
 parameters of the curvaton scenario~\cite{Nakayama:2009ce}. 
The present bound is $r_T<0.36$ (95\% CL)~\cite{Komatsu:2010fb}, 
 which is expected to be tightened as $r_T\simeq 10^{-1}$ from Planck satellite~\cite{Planck}
 and $r_T\simeq 10^{-3}$ 
 from DECi-hertz Interferometer Gravitational wave Observatory (DECIGO)~\cite{DECIGO} 
 and/or the Big Bang Observatory(BBO)~\cite{BBO}.

Curvaton scenarios have been often modeled
 by a scalar field $\sigma$ 
 with a quadratic potential $V= \frac{1}{2}m_{\sigma}^2 \sigma^2$.
In that case the non-Gaussianity has been studied with
 the sudden decay approximation~\cite{Bartolo:2003jx,Lyth:2005fi,Sasaki:2006kq},
 which shows good agreement with the full numerical approach~\cite{Malik:2006pm}.
Beyond the simplest model of the curvaton, there are various possibilities; 
 the inflaton perturbation may not be negligible~\cite{MixedIC},
 the curvaton can have different types of
 potential~\cite{Dimopoulos:2003ss,Huang:2008zj,Chingangbam:2009xi,Enqvist:2009zf} and
 there could be multiple curvaton fields~\cite{Choi:2007fya,Assadullahi:2007uw,Huang:2008rj}.
Significant effects on non-Gaussianity due to non-quadratic terms
 can be seen~\cite{Enqvist:2005pg,Enqvist:2008gk,Huang:2008bg,Enqvist:2009eq,Enqvist:2009ww,Byrnes:2010xd}.

So far, it has been assumed that the mass squared 
 at the origin of field is positive as above.
However, there is no reason that the true minimum is located at the origin.
Scalar fields have been often introduced 
 for spontaneous symmetry breaking in particle physics models.
A moduli field, which is a promising 
 candidate of curvaton~\cite{CurvatonMT},
or the Peccei-Quinn field, to solve the strong CP problem,
 usually has the large vacuum expectation value (VEV).

In this paper, we examine a curvaton model with a double well potential 
 where it develops nonvanishing VEV.
Since a curvaton needs to develop large expectation value during inflation,
 throughout this work,
 we assume that the potential is very flat and 
 the self-coupling constant is small enough.
Such a tiny self-coupling scalar field model has been studied 
 in the axion model to solve the isocurvature perturbation and domain wall
 problems~\cite{Linde:1991km,Kasuya:1996ns}.
Due to the flatness the curvaton has a large initial amplitude,
$\sigma_* \gtrsim v$, and almost stays there 
during inflation with negligible movement.  
After inflation the curvaton field starts to roll down into the minimum,
 oscillates and decay into radiation.
After the curvaton decay, its isocurvature perturbation is transferred to
 the adiabatic curvature perturbation in the radiation dominated plasma.

We take account of the inflaton perturbation as well as that of the curvaton.
Therefore the curvature perturbation shows generalised mixed
 inflaton-curvaton type~\cite{MixedIC}.
We consider the initial amplitude of the curvaton field is arbitrary
 but larger than the symmetry breaking scale.
In the opposite case where the field starts rolling around hill
 of the potential, interesting results have been shown~\cite{Kawasaki:2008mc}.

The paper is organized as follows.
After describing the model with a double well potential in section~\ref{model},
 we consider the cosmological evolution of the scalar field in
 section~\ref{evolution}. In section~\ref{perturbation} we estimate 
the non-Gaussianity and gravitational wave background of this model.
We summarize our results in section~\ref{summary}.

\section{A curvaton model with a double well potential}
\label{model}

We consider a real scalar curvaton model with a double well potential.
The Lagrangian density of the field is given by
\begin{equation}
 {\cal L} = -\frac{1}{2} (\partial \sigma)^2 -V(\sigma) ,
\end{equation}
\begin{equation}
 V(\sigma) = \frac{\lambda}{4} \left(\sigma^2 - v^2 \right)^2 ,
 \label{potential} 
\end{equation}
 with $\lambda$ and $v$ being respectively
 the self coupling constant and the VEV.
When the field has a nontrivial expectation value in the potential \eq{potential},
 the effective mass of it is expressed as
\begin{equation}
 V_{\sigma\sigma} =\lambda (3 \sigma^2 - v^2)
\end{equation}
and the mass at the true vacuum is given by
\begin{eqnarray}
 m_{\sigma}^2 = \left.V_{\sigma\sigma}\right|_{\sigma = \pm v} = 2 \lambda v^2.
\end{eqnarray}
The decay rate of $\sigma$ depends 
 on its interaction with light particles.
If $\sigma$ couples with a light fermion $\psi$ through a Yukawa interaction as 
 ${\cal L}_{\rm int} = y\bar{\psi}\sigma\psi$,
 the decay rate is roughly given by
\begin{equation}
 \Gamma_{\sigma} \simeq \frac{y^2}{8\pi} m_{\sigma} .
 \label{Gamma:YukawaType}
\end{equation}
If $\sigma$ does not directly couple with light particles,
 the decay rate would be expressed as
\begin{eqnarray}
 \Gamma_{\sigma} = C \frac{m_{\sigma}^3}{v^2} = C (2\lambda)^{3/2}v .
 \label{Gamma:SaxionType}
\end{eqnarray}
 with $C$ being a numerical coefficient of including 
 coupling constants and phase volume.
This kind of decay rate formula is realized, for instance, for 
 the radial direction of Peccei-Quinn field 
 in the hadronic (KSVZ) axion model~\cite{KSVZ}~\footnote{
 The corresponding curvaton scenario with Peccei-Quinn field in the extension 
of MSSM has been studied in Ref.~\cite{Dimopoulos:2003ii}.}.

\section{Cosmological evolution of $\sigma$}
\label{evolution}

By definition, the curvaton field is subdominant during inflation.
Its contribution can be important after inflation in the deep radiation
dominated era.
For this purpose, we consider the case that $\lambda$ is very tiny.
In such a case, the potential is very flat, 
 like chaotic inflation with quartic potential, 
 and hence the fields can develop a large expectation value during inflation.

The equation of motion for the homogeneous part of $\sigma$ is given by
\begin{eqnarray}
\ddot{\sigma}+3 H \dot{\sigma}+\lambda (\sigma^2 - v^2) \sigma = 0.
\end{eqnarray}
Before the inflaton decay,
 the energy density of the Universe is dominated by the inflaton whose
 equation of motion is given by
\begin{equation}
\ddot{\phi}+3 H \dot{\phi}+\frac{dV}{d\phi} = 0,
\end{equation}
 and then the Friedmann equation is
\begin{equation}
 3 M_P^2 H^2=\rho_\phi + \rho_\sigma,
\end{equation}
 with $\rho_\phi \gg \rho_\sigma$.
Here $M_P \simeq 2.4\times 10^{18}$ GeV denotes the reduced Planck mass.
After the inflaton decay, the Universe is dominated by the radiation generated
from the inflaton decay.,
Then the field equations, instead, are
\dis{
&\dot{\rho_r}+4H\rho_r=0,\\
&3 M_P^2 H^2=\rho_r + \rho_\sigma.
}
The curvaton field with a large expectation value during inflation almost
 stays there until 
 the Hubble parameter $H$ becomes comparable with the effective mass, i.e.
\begin{eqnarray}
 H_{\rm os}^2 \simeq \left.V_{\sigma\sigma}\right|_{\rm os}=\lambda ( 3 \sigma_{\rm os}^2 - v^2).
\end{eqnarray}
 From that time, the curvaton starts to oscillate with the initial amplitude $\sigma_{\rm os}$ 
and the energy density
\begin{equation}
\left. \rho_{\sigma} \right|_{\rm os} = \frac{\lambda}{4} (\sigma_{\rm os}^2 - v^2)^2.
\label{rhoos}
\end{equation}
When the initial amplitude of the curvaton is much larger than the location of
 the minimum as $\sigma_*\gg v$,
 the evolution of the curvaton is dominated by the
 quartic potential and the hill at the origin can be ignored during the oscillation.
In this case, the energy density of the curvaton decreased as $a(t)^{-4}$ after 
the oscillation starts.

After the oscillation amplitude decreases enough so that
 the field cannot go across the potential hill around the origin,
 the field can settle down at one of the two degenerate and distinct vacua.
Once the curvaton field find one of the minimum,
 the oscillation amplitude becomes less than of the order of $v$,
 and the energy density becomes smaller than
\begin{eqnarray}
\left. \rho_{\sigma} \right|_v
 = \frac{\lambda}{4}(\sigma_v^2-v^2)^2 ,
 \label{rhov}
\end{eqnarray}
 with $\sigma_v$ being the amplitude of the order of $v$ that denotes
 the transition from quartic oscillation to quadratic one occurs.
Note that $\sv$ is independent of $\sigma_*$.
The ratio of the energy densities given by Eqs.~(\ref{rhoos}) and (\ref{rhov})
 is scaled by $(a_v/a_{\rm os})^{-4}$, 
 because the quartic term initially dominates.
The precise value of $\sigma_v$ does not affect the differentiation by 
$\sigma_*$ but it may affect the energy density of the curvaton when it decay,
since the exact transition epoch between quadratic and quartic affect the
evolution of the curvaton energy density.

\begin{figure}[t] 
\begin{center}
  \epsfig{file=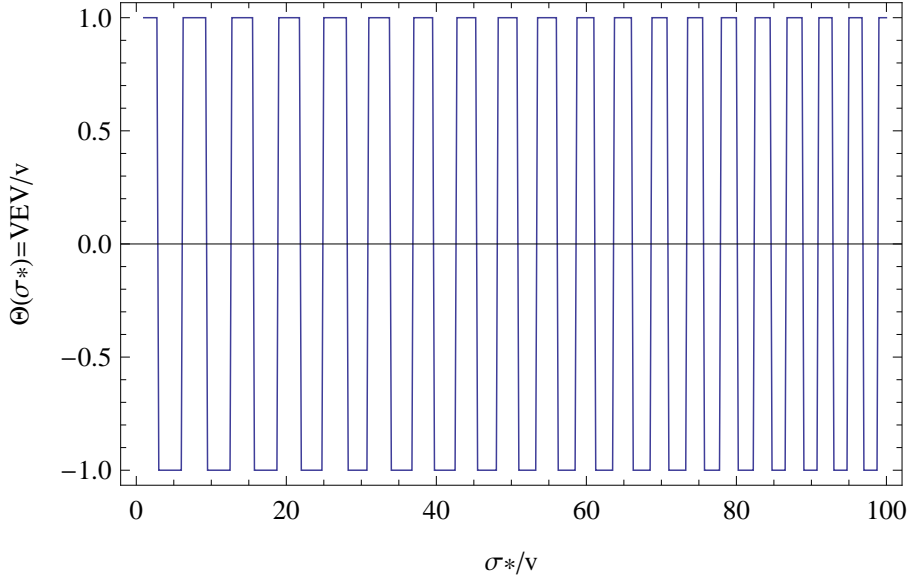,width=12cm}
\end{center}
 \caption{The position of VEV with varying initial amplitudes of the curvaton.
 The vertical axis is the sign of VEV, $\VEV{\sigma}/v$ 
 and the horizontal axis is $\sigma_*/v$. }
\label{fig:vev}
\end{figure}

After the curvaton find a minimum at $\langle \sigma \rangle
 = v$ or $\langle \sigma \rangle = -v$, 
 its energy density deceases as a pressureless matter $\propto a(t)^{-3}$
 since the quadratic potential dominates.
Which of the VEV would be realized depends on the initial field expectation value.
The dependence is shown as the function defined by 
$\Theta(\sigma_{\rm os})\equiv \VEV{\sigma}/v$ in figure~\ref{fig:vev}.

\begin{figure}[t] 
\begin{center}
  \epsfig{file=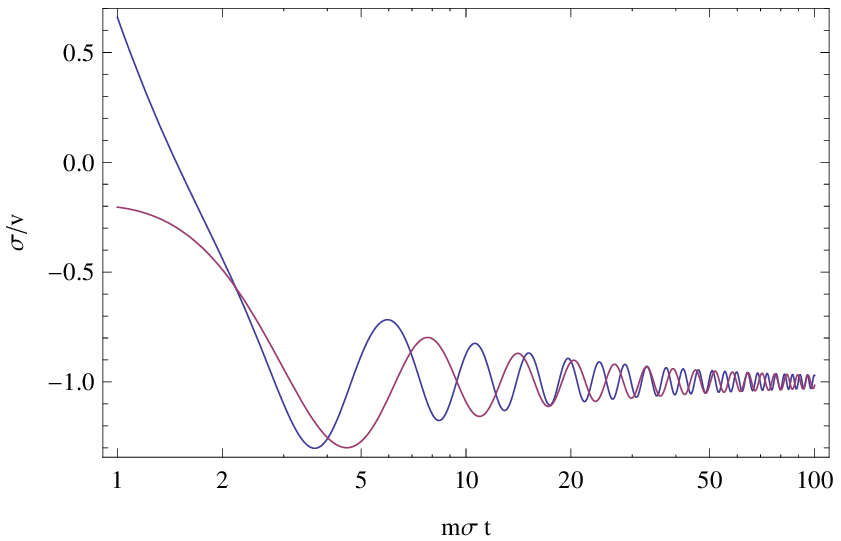,width=5cm}
  \epsfig{file=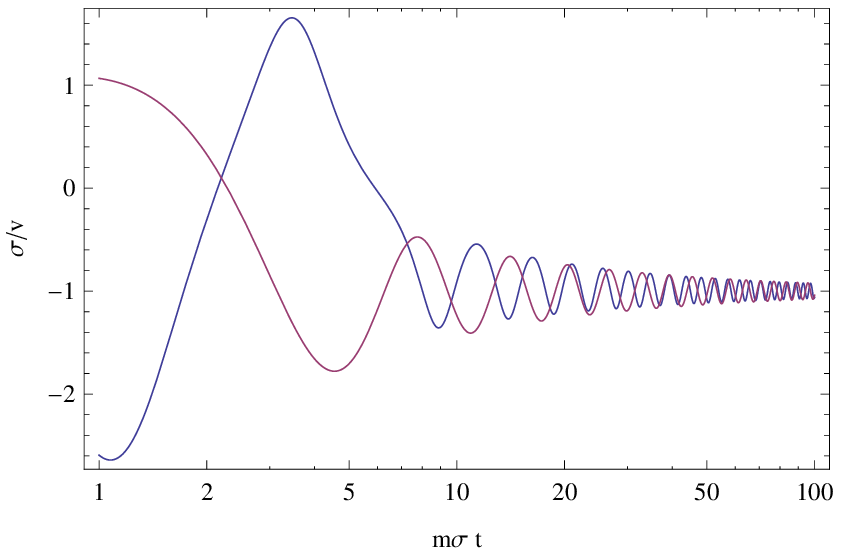,width=5cm}
  \epsfig{file=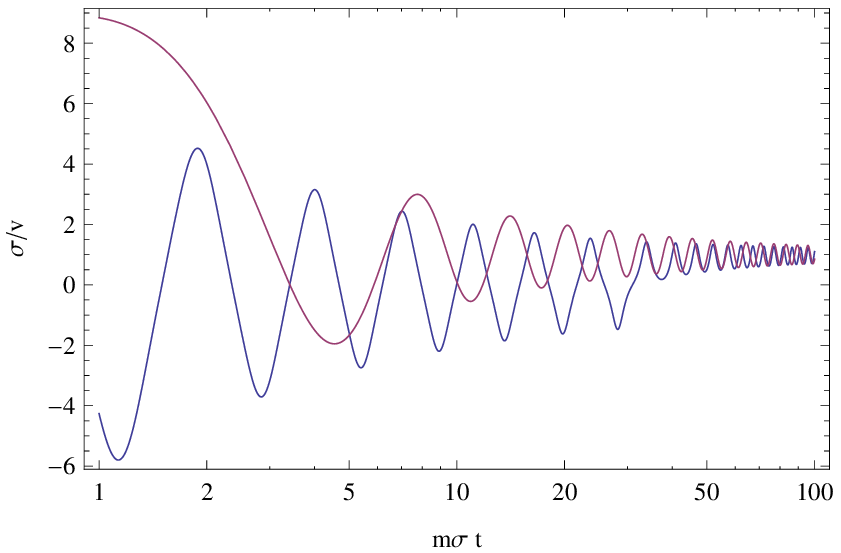,width=5cm}
\end{center}
 \caption{The comparison between analytic solutions (purple line) and 
 numerical solutions (blue line) 
 for various initial values $\sigma_{\rm os}= 4 v$ (left), $10 v$ (center),
 and $50 v$ (right). 
 The vertical axis is $\frac{\sigma}{v}$ and the horizontal axis is $m_{\sigma} t$.
 These figures show that analytic solutions well describe asymptotic evolution of the field.}
\label{fig:sols}
\end{figure}

At the late time $t\gg t_v$ in the deep oscillation period dominated 
by the quadratic potential,
the evolution of $\sigma$ can be well expressed as
\begin{equation}
 \sigma(t) \simeq v \Theta(\sigma_{\rm os}) + \frac{\sigma_{\rm 2os}}{(m_{\sigma} t)^{3/4}}\sin m_{\sigma}t .
\label{AnalyticalSolution}
\end{equation}
The amplitude of the oscillation $\sigma_{\rm 2os}$ can be estimated by
  using the simple scaling law between $t_{\rm os}$ and
  $t_v$~\cite{Enqvist:2009zf} 
for radiation dominated (high $\treh$) and oscillating inflaton dominated (low
$\treh$) at $H_{\rm os}$ respectively by 
\begin{equation}
\sigma_{\rm 2os}  \simeq
 \left\{   
\begin{array}{ll}
(\sv-v) \left(\frac{\left.\rho_\sigma\right|_{\rm os}}{\left.\rho_\sigma\right|_v}\right)^{3/8}
  \left(\frac{m_{\sigma} }
 {2 \sqrt{3\lambda  \sigma^2_{\rm os} }}\right)^{3/4},  
    & \textrm{for high $\treh$} , \\
(\sv-v) \left(\frac{\left.\rho_\sigma\right|_{\rm os}}{\left.\rho_\sigma\right|_v}\right)^{3/8}
\left(\frac{H_{R} }{2\sqrt{3\lambda \sigma^2_{\rm os} }}\right)^{1/4}
  \left(\frac{m_{\sigma} }
 {2 \sqrt{3\lambda \sigma^2_{\rm os} }}\right)^{3/4},  
    & \textrm{for low $\treh$} ,\\
\end{array}  
\right.
 \label{2Amplitude} 
\end{equation}
 with
\begin{eqnarray} 
\frac{1}{2 t_v} \simeq H_v = m_{\sigma} 
 \left(
 \frac{\sigma_{\rm 2os}}{\sigma_v-v} 
 \right)^{-4/3} ,
\label{tv} 
\end{eqnarray}
Here, the time of reheating can be approximated when the Hubble parameter
 is similar as the decay rate of inflation $\Gamma_{\phi}$ and the reheating
 temperature $\treh$ is estimated by
\begin{equation}
 \Gamma_{\phi}^2=H_R^2=\frac{1}{3\Mp^2}\frac{\pi^2}{30}g_*\treh^4.
\end{equation}
We consider $\Gamma_{\phi}$ or $\treh$ as a free parameter.
Equation~(\ref{2Amplitude}) has been sometime noted as $g$~\cite{Sasaki:2006kq} or 
 $\sigma_{\rm os}$~\cite{Enqvist:2009zf} in literature.
Figure \ref{fig:sols} shows the good agreement 
 between analytic approximated solutions \eq{AnalyticalSolution}
 and full numerical solutions.
Finally, the $\sigma$ field decays into radiation,
 when the Hubble parameter $H$ becomes comparable with its decay rate
 $H \simeq \Gamma_{\sigma}$.

When $v<\sigma_*< \sigma_v$, the curvaton field starts to oscillate 
initially in the potential dominated by quadratic term when $H^2\simeq m_\sigma^2$.
Therefore we find that 
\dis{
\sigma_{\rm 2os} \simeq (\sigma_* -
v\Theta)(m_\sigma/2H_{\rm os})^{3/4}.\label{lowsigmatwo}
}


Here we summarize the conditions for the $\sigma$ field to be a viable candidate 
for curvaton.
The curvaton is almost massless and its field value is frozen 
 during inflation. This is expressed by $V_{\sigma\sigma} \ll H_{\rm inf}^2$
 and rewritten as
\begin{eqnarray}
\textrm{(I)}\qquad 3 \lambda (\sstar^2 -v^2)\ll H_{\rm inf}^2.
\label{FieldFrozenCondition}
\end{eqnarray}
The curvaton energy density is subdominant compared
 with that of the inflaton during inflation, which is expressed as
\begin{eqnarray}
 \frac{\lambda}{4}(\sstar^2 -v^2)^2 \ll 3 M_P^2 H_{\rm inf}^2.
\label{SubdominantCondition}
\end{eqnarray}
This condition is automatically satisfied from \eq{FieldFrozenCondition}
 when $\sstar < \Mp$.
Whether $\sigma$ dominates the energy density of the Universe
 at the moment of $\sigma$ decay
 depends on the reheating temperature after inflation $T_R$ determined 
 by the inflaton decay rate $\Gamma_{\phi}$.
We will pursue the details for this in the following subsections.
During preheating, the symmetry might be restored and the topological 
defects could be formed due to the parametric
resonance and the large fluctuation of $\VEV{\delta \sigma^2}\gg v^2$
~\cite{Kofman:1995fi}.
This problem can be avoided if the dynamics is pure classical and 
the initial fluctuations $\delta\sigma/\sigma$ is less than the change of the
amplitude
of $\sigma$ per one oscillation when the curvaton field settles down to one
of two minima of the potential~\cite{Kasuya:1996ns}.
This condition is easily written down as
\dis{
\frac{\delta \sigma}{\sigma}\simeq \frac{H_{\rm inf}/2\pi}{\sigma_{\rm os}} <
\frac{\Delta A}{A} \sim \frac{H_c}{\omega}\sim
\frac{\sqrt{\lambda}v^2/\sigma_{\rm os}}{\sqrt{\lambda}v},
}
where we have used the fact that the curvaton oscillation is dominated by
quartic term between the end of inflation and the critical point
and thus $\sigma_{\rm os}^4/\sv^4=H_{\rm os}^2/H_c^2$.
Therefore there is no domain wall problem for a large VEV satisfying
\dis{
\textrm{(II)} \qquad \frac{H_{\rm inf}}{2\pi} < v.
}
In this paper, we consider this large symmetry breaking scale.

\subsection{A high reheating temperature case}

First, we consider the case that the reheating after inflation 
 is completed when $\sigma$ starts to oscillate, which means that the Hubble
 parameter at reheating $H_R$ is larger than that at the beginning of the oscillation.
This condition of high reheating temperature corresponds
 to $H_R > H_{\rm os}$ or
\begin{equation}
 \treh > \left[\left(\frac{\pi^2}{30}g_*\right)^{-1}3\Mp^2
  \lambda(3 \sigma_{\rm os}^2-v^2)^2  \right]^{1/4}.
\label{highTR}
\end{equation}

The energy density of radiation produced by the inflaton decay at $H_{\rm os}$ is 
\begin{equation}
\left. \rho_r \right|_{\rm os} = 3 M_P^2 H_{\rm os}^2.
\end{equation}
The energy density of the curvaton and radiation from the inflaton decay 
 at $H \simeq \Gamma_{\sigma}$ are given by
\begin{eqnarray}
 \left. \rho_{\sigma} \right|_{\Gamma_{\sigma}}
  \simeq
  \frac{\lambda}{4}v^4 \left(\frac{a_v}{a_{\Gamma_{\sigma}}}\right)^3 ,
\end{eqnarray}
and
\begin{eqnarray}
 \rho_r
  = 3 M_P^2 H_{\rm os}^2 \left(\frac{a_{\rm os}}{a_{\Gamma_{\sigma}}}\right)^4 .
\end{eqnarray}
The $\sigma$ to radiation ratio is evaluated as
\beqa{
r \equiv \frac{\rho_{\sigma}}{\rho_r}
  &=& \frac{\lambda v^4}{12 M_P^2 H_{\rm os}^2} 
  \left(\frac{a_v}{a_{\Gamma_{\sigma}}}\right)^3
  \left(\frac{a_{\Gamma_{\sigma}}}{a_{\rm os}}\right)^{4} \nonumber \\ 
  &=& \frac{v^{1/2} \sigma_{\rm os}^{3/2}}{36 M_P^2}
  \left(\frac{3\lambda v^2}{\Gamma_\sigma^2}\right)^{1/4}
\label{PQratio}
}
 for the radiation dominated Universe. 

For the case of $v < \sigma_{\rm os} < \sigma_v $,
 we obtain
\begin{eqnarray}
r
 = \frac{\lambda^{1/4} (\sigma^2_{\rm os} -v^2)^2}{12 M_P^2 (3 \sigma_{\rm os}^2 - v^2)^{3/4}}
  \frac{1}{\Gamma_\sigma^{1/2} }
\end{eqnarray}
 for the radiation dominated Universe.

\subsection{A low reheating temperature case}

Next, we consider the case that the inflaton still oscillates
 (we assume the quadratic oscillation) around the minimum and
 the reheating is not completed yet when $\sigma$ starts to oscillate, 
 $H_R < H_{\rm os}$, which is the opposite condition of \eq{highTR}.

The energy density of the inflaton $\phi$ at $H_{\rm os}$ is 
\begin{equation}
\left. \rho_{\phi} \right|_{\rm os} = 3 M_P^2 H_{\rm os}^2.
\end{equation}
At a late time, the energy density of the curvaton and 
 radiation from the inflaton decay at $H \simeq \Gamma_{\sigma}$ are given by
\begin{eqnarray}
 \left. \rho_{\sigma} \right|_{\Gamma_{\sigma}}
  \simeq
  \frac{\lambda}{4}v^4 \left(\frac{a_v}{a_{\Gamma_{\sigma}}}\right)^3 ,
\end{eqnarray}
and
\begin{eqnarray}
 \rho_r
  = \frac{\pi^2}{30}g_* T_R^4 
  \left(\frac{a_{\rm R}}{a_{\Gamma_{\sigma}}}\right)^4 .
\end{eqnarray}
The energy density ratio of $\sigma$ to radiation at late time $H \simeq \Gamma_{\sigma}$
 is evaluated as
\begin{equation}
r = \frac{v \sigma_{\rm os}}{36 M_P^2}
  \left(\frac{\pi^2 g_* T_R^4}{90 M_P^2 \Gamma_\sigma^2}\right)^{1/4}
\end{equation}

For the case of $v < \sigma_{\rm os} < \sigma_v $,
 we obtain
\begin{eqnarray}
r
 = \frac{ (\sigma^2_{\rm os} -v^2)^2}{12 M_P^2 (3 \sigma_{\rm os}^2 - v^2)}
  \left(\frac{\pi^2 g_* T_R^4}{90 M_P^2 \Gamma_\sigma^2}\right)^{1/4}.
\end{eqnarray}

\section{Power spectrum and non-Gaussianity}
\label{perturbation}

The curvaton is light during inflation with \eq{FieldFrozenCondition}
 and thus has a Gaussian quantum fluctuation with the amplitude $\delta \sigma_* \simeq H_{*}/(2\pi)$.
The curvaton field value at the onset of its quadratic oscillation
 is some function of that at the onset of quartic oscillation
 $\sigma_{\rm 2os} = \sigma_{\rm 2os}(\sigma_{\rm os})$.
In addition we assume the field value at $t_{\rm os}$ is same as that at horizon exit $t_*$,
\begin{equation}
 \sigma_{\rm os}(\sigma_*)=\sigma_* .
 \label{sstar}
\end{equation}
Thus we can be expand $\sigma_{\rm 2os}$ around
 the homogeneous part $\sigma_{\rm 2os}$,
\begin{eqnarray}
\sigma_{\rm 2os}(t,x) = \sigma_{\rm 2os}(t) + \sigma_{\rm 2os}' \delta\sigma_* + \frac{1}{2} \sigma_{\rm 2os}'' (\delta\sigma_*)^2 
 + \frac{1}{6} \sigma_{\rm 2os}^{'''}(\delta\sigma_*) ^3 +\ldots,
\label{sigmaexpand}
\end{eqnarray}
 where the prime denotes the derivative with respect to $\sigma_*$. 

The curvature perturbation due to the curvaton density fluctuation is 
 easily calculated using $\delta N$
 formalism~\cite{Lyth:2005fi,Sasaki:2006kq}.
The nonlinear curvature perturbation of the curvaton field 
 on the uniform curvaton density hypersurface is given
 by~\cite{Lyth:2004gb,Sasaki:2006kq,Langlois:2008vk}, 
\dis{
\zeta_\sigma
 = \delta N +\frac13\int_{\rho_0(t)}^{\rho(t,{\bf x})}\frac{d\tilde{\rho}}{\tilde{\rho}+\tilde{p}},
}
 where $\delta N$ is the perturbed expansion, $\tilde{\rho}$ and $\tilde{p}$
 are the local density and pressure of the curvaton respectively.
For the oscillating curvaton field in the expanding Universe, the pressure and
 the energy density is related by $p=w\rho$ with $w=0$ 
 when quadratic term dominates and $w=1/3$ when quartic term dominates.
In our case, the curvaton energy density at late times ($t\gg t_v$)
 is given by
\dis{
\rho_\sigma(t,x) \simeq \frac{m_\sigma^2 \sigma_{\rm 2os}^2(t,x)}{2(mt)^{3/2}},
\label{rhooscillation}
}
and it can be expanded around the background value $\sigma_{\rm 2os}(t)$
 using \eq{sigmaexpand}.

From this, we can find the curvature perturbation of the curvaton field as
\begin{eqnarray}
\zeta_{\sigma}
 = \zeta_{\sigma 1}+ \frac{1}{2} \zeta_{\sigma 2}  + \frac{1}{6} \zeta_{\sigma 3}+\ldots,
\label{zetasA}
\end{eqnarray}
 where each terms are expressed as~\cite{Huang:2008zj}
\begin{eqnarray}
\zeta_{\sigma 1}
 &=& \frac{2\sigma_{\rm 2os}'}{3\sigma_{\rm 2os}}\delta\sigma_*,\label{zetas1}\\
\zeta_{\sigma 2}
 &=& -\frac{3}{2}\left(1-\frac{\sigma_{\rm 2os}\sigma_{\rm 2os}''}{\sigma_{\rm
    2os}^{'2}} \right)\zeta_{\sigma 1}^2\equiv A_2\zeta_{\sigma 1}^2,\label{A2}\\
\zeta_{\sigma 3}
 &=& \frac94\left(2-3\frac{\sigma_{\rm 2os}\sigma_{\rm 2os}''}{\sigma_{\rm
    2os}^{'2}}+ \frac{\sigma_{\rm 2os}^2\sigma_{\rm 2os}'''}{\sigma_{\rm 2os}^{'3}} \right)\zeta_{\sigma 1}^3\equiv A_3\zeta_{\sigma 1}^3.
\end{eqnarray}
On this uniform density surface 
 at the curvaton decay time $t_D$ or $H=\Gamma_{\sigma}$, we have
\begin{eqnarray}
 \rho_r(t_D) e^{4(\zeta_r -\zeta)} + \rho_\sigma(t_D) e^{
   3(\zeta_{\sigma} -\zeta)} = \rho_{\rm tot}(t_D),
\end{eqnarray}
 with the radiation perturbation $\zeta_r$ originated from the inflaton $\phi$.
Then, the curvature perturbation after the curvaton decay can be expressed as
\begin{equation}
 \zeta = \zeta_1+\frac{1}{2}\zeta_2+\frac{1}{6}\zeta_3+\ldots,
\label{zetaA}
\end{equation} 
where
\dis{
\zeta_1=&(1-R)\zeta_{r1} +R\zeta_{\sigma 1},\\
\zeta_2=&(1-R)\zeta_{r2} +R\zeta_{\sigma 2}
+R(1-R)(3+R)\left(\zeta_{r1}-\zeta_{\sigma 1}\right)^2,\\
\zeta_3=&(1-R)\zeta_{r3} +R\zeta_{\sigma 3}
+3R(1-R)(3+R)\left(\zeta_{r1}-\zeta_{\sigma
  1}\right)\left(\zeta_{r2}-\zeta_{\sigma 2}\right)\\
+& R(1-R)(3+R)(-3+4 R +3 R^2)\left(\zeta_{r1}-\zeta_{\sigma 1}\right)^3,
}
and
\dis{
R\equiv \frac{3\rho_\sigma}{4\rho_r + 3\rho_\sigma}, 
}
 at $t=t_D$. 
Furthermore it is natural to assume $\zeta_r\ll\zeta_\sigma$ and
$\zeta_r$ is almost Gaussian so that only $\zeta_{r1}$ is non-negligible in the expansion of $\zeta_r$.
Thus in the above we can approximate $\zeta_{r1}-\zeta_{\sigma 1}\simeq
-\zeta_{\sigma 1} $ and $\zeta_{r2}-\zeta_{\sigma 2}\simeq -\zeta_{\sigma 2}$.

The power spectrum is obtained as
\begin{equation}
{\mathcal P}_{\zeta} = (1-R)^2 {\cal P}_{\zeta_{r}} + R^2 {\mathcal
  P}_{\zeta_{\sigma}}.
\label{Pzeta}
\end{equation}
 by using Eqs.~(\ref{zetasA}) and (\ref{zetaA}).
In \eq{Pzeta}, the spectrum of radiation and the curvaton is given by
\dis{
{\mathcal P}_{\zeta_r} =& \left(\frac{H_*^2}{2 \pi |\dot{\phi}|}\right)^2
 = \frac{H_*^2}{8 \pi^2 \epsilon M_P^2},\\
{\mathcal P}_{\zeta_\sigma}=&\frac{H_*^2}{4\pi^2} 
\left(\frac{2\sigma_{\rm 2 os}' }{3\sigma_{\rm 2 os}}\right)^2
\label{Pzetaeach}
}
with
\begin{eqnarray}
 \epsilon \equiv \frac{M_P^2}{2}\left(\frac{V_{\phi}}{V}\right)^2
  \simeq \frac{|\dot{H}|}{H^2} .
\end{eqnarray}
We defined $\tr$ as the ratio of the contribution to the linear perturbation
of $\zeta$ from the curvaton to that from the inflaton, i.e.
\dis{
\tr\equiv \frac{\RD^2{\mathcal P}_{\zeta_\sigma}}{(1-\RD)^2{\mathcal
    P}_{\zeta_r}}=\frac{\RD^2}{(1-\RD)^2}2\epsilon 
 \left(\frac{2\sigma_{\rm 2 os}'}{3\sigma_{\rm 2 os}}\right)^2 M_P^2.
\label{tr}
}
In the limit of $\tr \rightarrow \infty$, \eq{Pzeta} becomes that of simple
curvaton scenario neglecting the inflaton contribution and in the opposite limit
$\tr\rightarrow 0$ the power spectrum has dominant contribution from that of the inflaton.
Although for both cases the observable non-Gaussianity is
possible~\cite{Byrnes:2010em}, the constraint from tensor-to-scalar ratio
disfavors the small $\tr$ region as we will see later.

 The scalar spectral index is given by
\begin{eqnarray}
 n_s = 1- 2 \epsilon + \frac{2 V_{\sigma\sigma}}{3 H_*^2} ,
\label{ns}
\end{eqnarray}
 and we find $2 \epsilon \simeq 0.04$ from the WMAP~\cite{Komatsu:2010fb} 
 data, $n_s \simeq 0.96$.
With this value of $\epsilon$, \eq{Pzeta} and \eq{Pzetaeach} relate $H_*$
 with the primordial power spectrum for the given initial value of $\sigma_{\rm os}$.
Therefore we can obtain the each contribution from the curvaton and the inflaton
separately and thus $\tilde{r}$ in our scenario.

The tensor perturbation (gravitational wave) is also generated
during inflation~\cite{GWbackground}.
The tensor to scalar ratio $r_T$ is given by
\beqa{
r_T&=& \frac{P_T}{P_{\zeta}}= \frac{16\epsilon}{(1-R)^2(1+\tr)}
\label{TensorConstraint}
}
If the amplitude is large enough, 
 the gravitational wave background is detectable 
 through the measurement of the B-mode polarization 
 in the cosmic microwave background(CMB) anisotropy by Planck~\cite{Planck}
 as well as the direct detection by future interferometers such as
 DECIGO~\cite{DECIGO}.

We obtain the nonlinearity parameters 
\begin{eqnarray}
f_{NL} &=& 
 \frac{5}{6}\frac{\tilde{r}^2}{(1+\tilde{r})^2}\left[ \frac{3+A_2}{R}
   -2-R  \right], \label{fnl}
\end{eqnarray}
\begin{eqnarray}
g_{NL} &=&
\frac{25}{54}\frac{\tilde{r}^3}{(1+\tilde{r})^3}\left[\frac{9+9A_2+A_3}{R^2}
  -\frac{18+6A_2}{R}-4-3A_2 +10R +3R^2 \right].\label{gnl}
\end{eqnarray}
Here we have assumed that $\zeta_r$ is Gaussian so that only
  $\zeta_{r1}$ is non-zero and  $\zeta_{r1}\ll\zeta_{\sigma 1}$ which is true in the curvaton scenario when $\sstar\ll \Mp$.

The WMAP data on the power spectrum of scalar and tensor perturbation 
\dis{
{\cal P}_{\zeta} \simeq 2.4\times10^{-9}, \qquad r_T < 0.36,\label{Pzetawmap}
} 
as well as the local type non-linearity parameter
\dis{
-10 < \fnl < 74,
}
 constrain possible values of $H_*$, $\sigma_*$ and $R$.

\subsection{small initial expectation value}

For a small initial amplitude of the curvaton field $v<\sigma_*<\sv$,
the oscillation starts when the quadratic term dominates.
In this case, from \eq{lowsigmatwo} with $\sos=\sigma_*$,
we find that
\dis{
\stwo'=&\frac{(m_\sigma/2\sqrt{\lambda})^{3/4}}{(3\sos^2-v^2)^{3/8}}\left(1-\frac94\frac{\sos(\sos-v)}{(3\sos^2-v^2)}\right)\\
\stwo''=&\frac{(m_\sigma/2\sqrt{\lambda})^{3/4}}{(3\sos^2-v^2)^{11/8}}\left(-\frac92\sos
-\frac94(\sos-v)+\frac{297}{16}\frac{(\sos-v)\sos^2}{(3\sos^2-v^2)}\right),\\
\stwo'''=& \frac{(m_\sigma/2\sqrt{\lambda})^{3/4}}{(3\sos^2-v^2)^{11/8}}\left(-\frac{27}{4}+\frac{891}{16}\frac{\sos^2+(\sos-v)\sos}{(3\sos^2-v^2)}
-\frac{16929}{64}\frac{(\sos-v)\sos^3}{(3\sos^2-v^2)^2}\right).
}
The curvature perturbation of the curvaton $\zeta_\sigma$ and the corresponding
non-linearity parameters are evaluated from this.
Note that there are additional factors coming from the
dependence on $H_{\rm 2 os}$ compared to the simple curvaton model with
quadratic potential.
One thing to note is that $A_2$, defined in \eq{A2}, becomes negative when $\sigma_{\rm os} \gtrsim
1.8v$, which changes the sign of $\fnl$ in \eq{fnl} for small $\RD$.
The non-trivial behavior from this is shown in the lower part of Figs.~\ref{fig:fnlTR12} and~\ref{fig:fnlTR14}.

For the pure quadratic potential limit $\sigma_{\rm os} \simeq v$,
 the expression of $\zeta_\sigma$ is reduced to
\dis{
\zeta_\sigma=\frac23 \frac{\delta \sigma_*}{\sigma_*-v}- 
  \frac13 \bfrac{\delta\sigma_*}{\sigma_*-v}^2 
  +\frac29\bfrac{\delta\sigma_*}{\sigma_*-v}^3 .
}
The density perturbation of radiation after the curvaton decay is
\dis{
\zeta=(1-\RD) \zeta_r+ \frac{2\RD}{3} \bfrac{\delta \sigma_*}{\sigma_*-v}
+&\frac29\left(\frac{3}{2\RD}-2-\RD\right)R^2\bfrac{\delta\sigma_*}{\sigma_*-v}^2\\
+&\frac{4}{81}\left(-\frac{9}{\RD}+\frac12+10\RD+3\RD^2\right)R^3\bfrac{\delta\sigma_*}{\sigma_*-v}^3+\ldots. \label{zeta2}
}
The nonlinearity parameters are given for $v<\sigma_*<\sv$,
 from Eqs.~(\ref{fnl}), (\ref{gnl}) and (\ref{zetasA}), 
\dis{
\fnl &= \bfrac{\tr}{1+\tr}^2\frac56\left(\frac{3}{2\RD}-2-\RD \right)>-\frac32\\
\gnl &= \bfrac{\tr}{1+\tr}^3\frac{25}{54}\left( -\frac{9}{\RD} +\frac12 +10\RD+3\RD^2 \right).\label{nonG2}
} 
The sizable large $\fnl \sim 100$ is obtained with a small ratio $\RD \sim 10^{-2}$.

\subsection{large initial expectation value}

\subsubsection{evolution of perturbations}

\begin{figure}[t]  
\begin{center}
  \epsfig{file=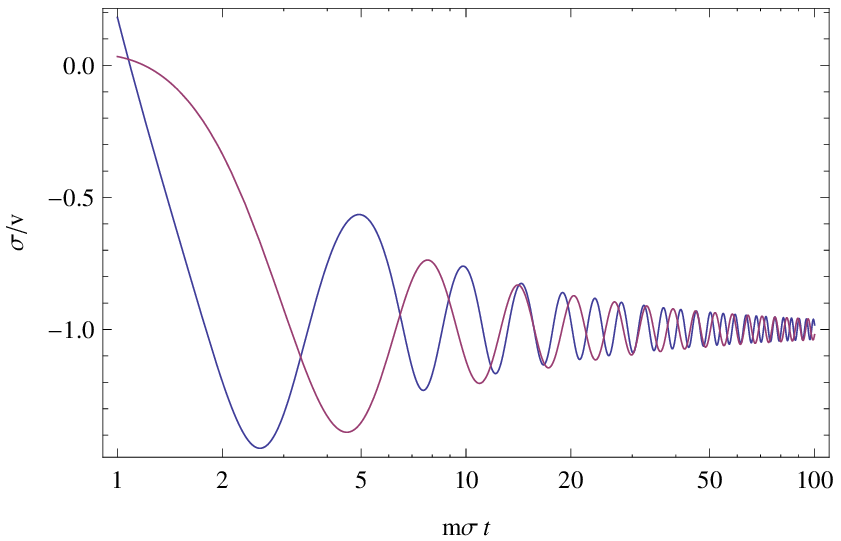,width=5cm}
  \epsfig{file=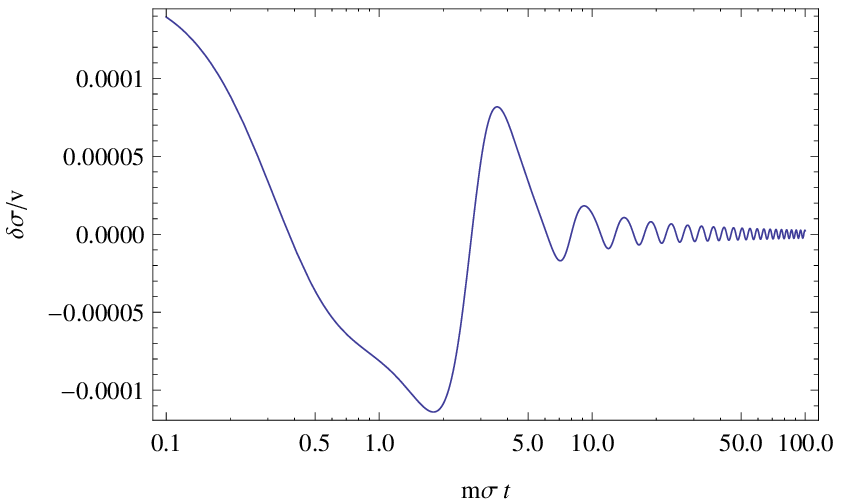,width=5cm}
  \epsfig{file=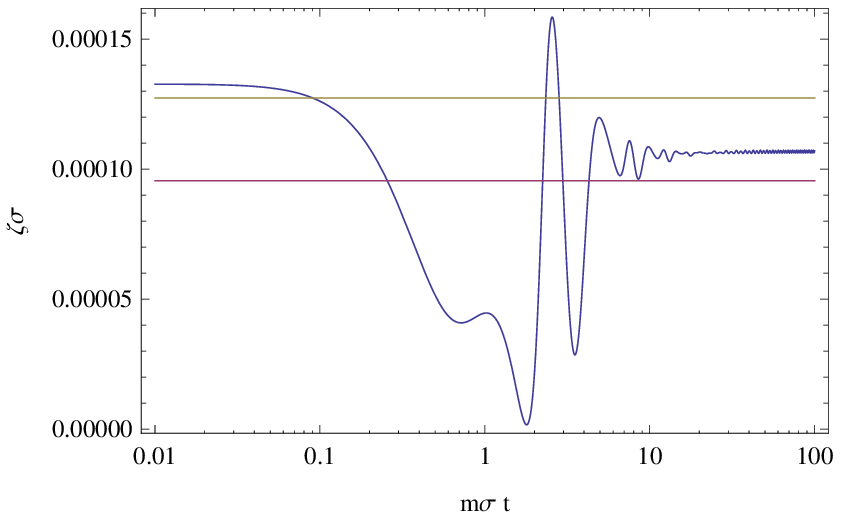,width=5cm}
\end{center}
\begin{center}
  \epsfig{file=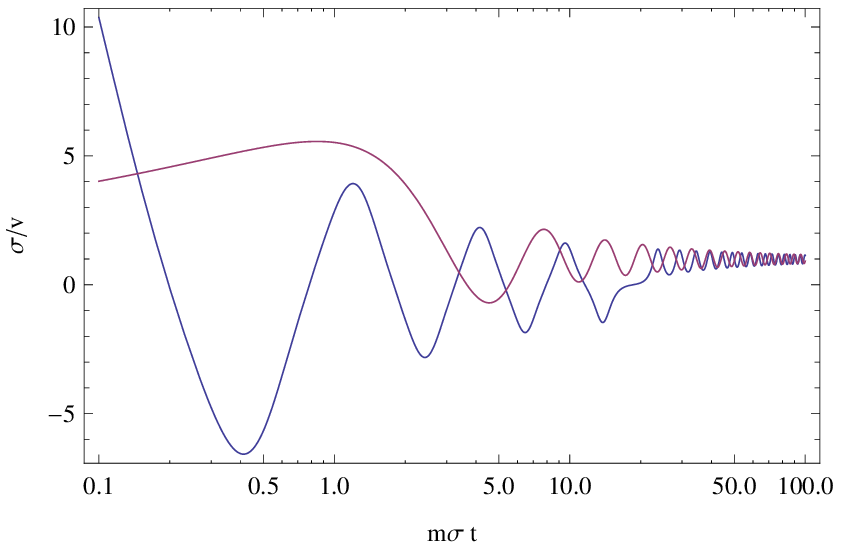,width=5cm}
  \epsfig{file=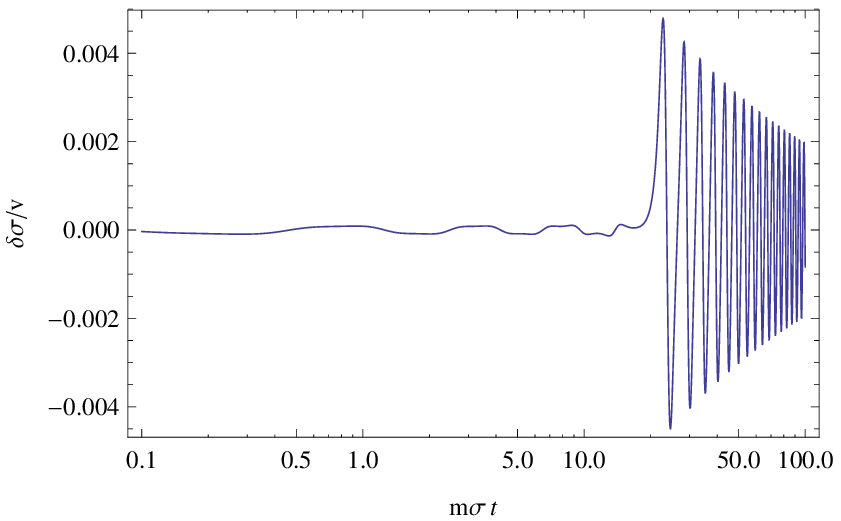,width=5cm}
  \epsfig{file=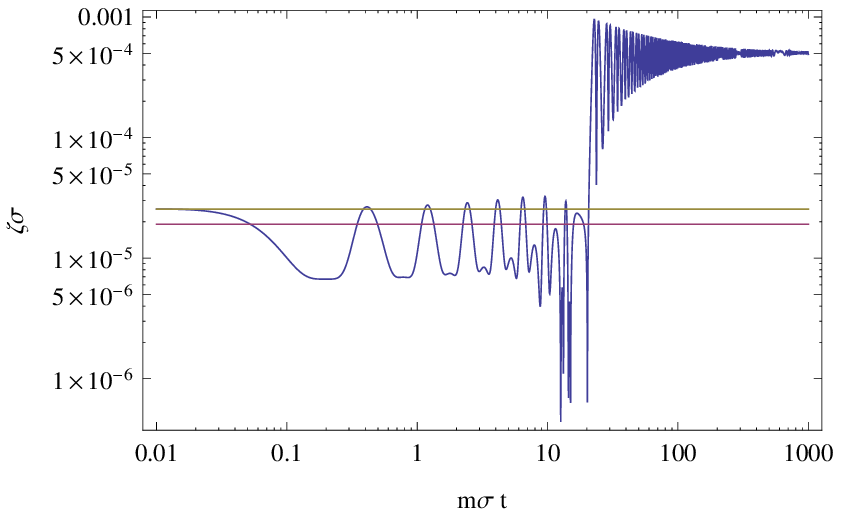,width=5cm}
\end{center}
\begin{center}
  \epsfig{file=in50.eps,width=5cm}
  \epsfig{file=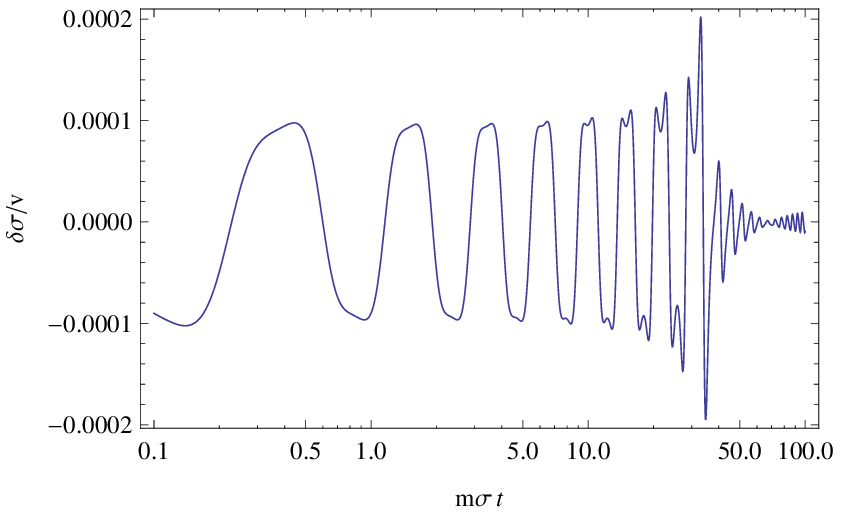,width=5cm}
  \epsfig{file=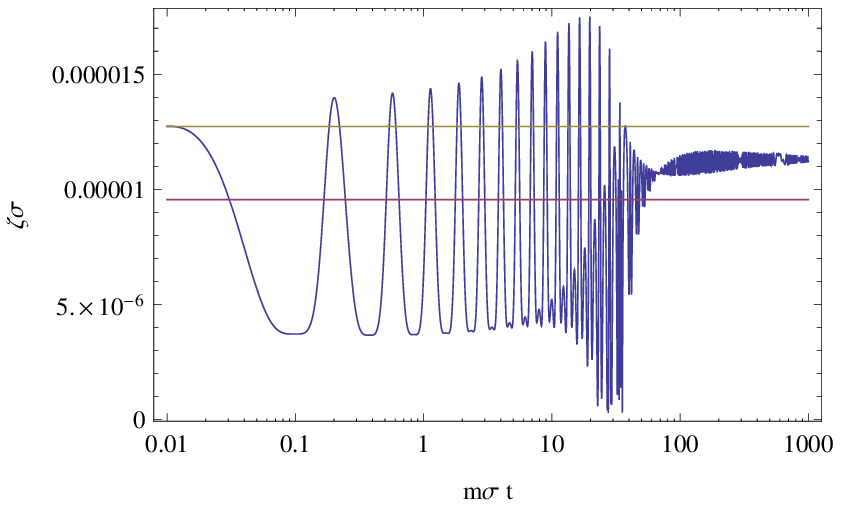,width=5cm}
\end{center}
 \caption{
The evolution of $\sigma$ (left), $\delta\sigma$ (center), and
$\zeta_{\sigma}$ (right).
The upper (middle, lower) row
 corresponds to the results for $\sigma_{\rm os} = 5\, (25, 50)\, v$.
Notice that the vertical axis of the right-middle figure is logarithmic scale.
Here, we assume $\delta\sigma_* = H_*/(2\pi)$ with $H_* = 10^{-3} v$.
The green (purple) line in the right figures
 expresses the analytic formula of $\zeta_{\sigma}$
 at $t_{\rm os}$($t_v$) without including the amplification effect.
These show that 
 the error of the analytic formula is just about ${\cal O} (10) \% $ 
 unless the tachyonic instability is induced as for $\sigma_{\rm os} = 5v, 50v$.
}
\label{fig:zeta}
\end{figure}

Next, we consider a large initial amplitude of the curvaton field $\sigma_* \gg v$.
At the early stage of oscillation, the field evolution is
 due to the quartic potential and highly nonlinear.
For our model, the corresponding quantity $\sigma_{\rm 2os}$ 
 is analytically related to $\sigma_*$ using Eqs.~(\ref{2Amplitude}) and (\ref{sstar}),
 ignoring the $\Theta$ part.
Then, we obtain
\begin{eqnarray}
&& \frac{1}{\sigma_{\rm 2os}}\frac{d \sigma_{\rm 2os}}{d \sigma_*}
 \simeq \frac{3}{4\sigma_*}, \nonumber \\ 
&& \frac{\sigma_{\rm 2os}\sigma''_{\rm 2os}}{(\sigma'_{\rm 2os})^2}
 \simeq -\frac{1}{3}, \label{dsigmahigh}  \\
&& \frac{\sigma^2_{\rm 2os}\sigma'''_{\rm 2os}}{(\sigma'_{\rm 2os})^3}
 \simeq \frac{5}{9}, \nonumber
\end{eqnarray}
 for a high reheating temperature case, and 
\begin{eqnarray}
&& \frac{1}{\sigma_{\rm 2os}}\frac{d \sigma_{\rm 2os}}{d \sigma_*}
 \simeq \frac{1}{2\sigma_*}, \nonumber \\   
&& \frac{\sigma_{\rm 2os}\sigma''_{\rm 2os}}{(\sigma'_{\rm 2os})^2}
 \simeq -1,  \label{dsigmalow} \\
&& \frac{\sigma^2_{\rm 2os}\sigma'''_{\rm 2os}}{(\sigma'_{\rm 2os})^3}
 \simeq 3, \nonumber 
\end{eqnarray}
 for a low reheating temperature case.
The prime denotes the derivative with respect to $\sigma_*$.
In addition, concerning with the $\Theta$ part, because of the high nonlinearity,
 the fluctuation $\delta\sigma$ also undergoes nontrivial evolution.
The equation of motion for $\delta\sigma$ of superhorizon scale ($k \ll H a$) is given by
\begin{eqnarray}
\ddot{\delta\sigma}+3 H \dot{\delta\sigma}+\lambda (3\sigma^2 - v^2) \delta\sigma = 0.
\end{eqnarray}
During $\sigma \simeq 0$, $\delta\sigma$ has effectively the negative mass. 
This tachyonic instability 
 leads to significant amplification of the fluctuation $\delta\sigma$,
 in some cases 
 that the initial value $\sigma_{\rm os}$ corresponds to 
 the transition of the VEV from $-v$ to $v$ 
 in Fig.~\ref{fig:vev}.~\cite{Dimopoulos:2003ss}.
Figure~\ref{fig:zeta} shows the amplification and evolution of 
 the field fluctuation $\delta\sigma$ and the density $\zeta_{\sigma}$ 
 for some $\sigma_{\rm os}$s.
For cases in which the field $\sigma$ stays near the origin longer,
 the amplification is sizable and, with Eq.~(\ref{tv}), roughly estimated as
\begin{equation}
T \equiv \frac{\delta\sigma_{\rm 2os}}{\delta\sigma_*} \sim e^{\sqrt{\lambda} v \Delta t}
 < e^{\sqrt{\lambda} v t_v} \sim e^{\frac{\sigma_{\rm os}}{v}} 
\label{Amplify}
\end{equation}
 for a high reheating case, as seen 
 in the middle row in Fig.~\ref{fig:zeta} for $\sigma_{\rm os} = 25 v$.
Here $\Delta t$ denotes the period during the tachyonic instability works. 
The final fluctuation after quadratic oscillation starts is,
 with the amplification factor $T$, given by 
\begin{eqnarray}
\left.\delta\sigma\right|_{\rm 2os} = T \frac{d \sigma_{\rm 2os}}{d \sigma_*} \delta\sigma_* .
\end{eqnarray}
However, this enhancement occurs only for limited conditions of $\sigma_{\rm os}$
 near the VEV transition intial expectation value.
Thus, from now on, we consider cases without this enhancement and
 these enhanced modes will be studied in future works.
Then, we obtain the curvature perturbation of the curvaton field
\dis{
\left.\zeta_\sigma\right|_{t \gg t_v} &=\frac12 \frac{\delta \sigma_*}{\sigma_*}- 
  \frac14 \bfrac{\delta\sigma_*}{\sigma_*}^2 
  +\frac16 \bfrac{\delta\sigma_*}{\sigma_*}^3,\qquad \textrm{for high $\treh$},\\
\left.\zeta_\sigma\right|_{t \gg t_v} &=\frac13 \frac{\delta \sigma_*}{\sigma_*}- 
  \frac16 \bfrac{\delta\sigma_*}{\sigma_*}^2 
  +\frac19\bfrac{\delta\sigma_*}{\sigma_*}^3,\qquad \textrm{for low $\treh$}.
\label{zetasigmaboth}
}
This is conserved until the curvaton decay
 for $t \gg t_v$, as seen in figure.~\ref{fig:zeta}.

\subsubsection{observables}

Since the total curvature perturbation of radiation after the curvaton decay is
conserved, it can be calculated at the time of the curvaton decay.
For the case of double well potential, i.e. initially quartic term dominates
and the mass term becomes dominant before the curvaton decays, 
the primordial curvature perturbation is obtained,
 from Eqs.~(\ref{zetaA}) and (\ref{zetasigmaboth}) with $\sigma_\os=\sigma_*$, as
\dis{
\zeta=(1-\RD) \zeta_r+& \frac{\RD}{2} \bfrac{\delta \sigma_*}{\sigma_*}
+\frac18\left(\frac{1}{\RD}-2-\RD\right)R^2\bfrac{\delta\sigma_*}{\sigma_*}^2\\
+&\frac{1}{48}\left(-\frac{1}{\RD^2}-\frac{6}{\RD}+2+10\RD+3\RD^2
\right)R^3\bfrac{\delta\sigma_*}{\sigma_*}^3, \qquad\textrm{for high
  $\treh$},\label{zetahighTR}
}
and 
\dis{
\zeta=(1-\RD) \zeta_r+& \frac{\RD}{3} \bfrac{\delta \sigma_*}{\sigma_*}
+\frac{1}{18}\left(-2-\RD\right)R^2\bfrac{\delta\sigma_*}{\sigma_*}^2\\
+&\frac{1}{162}\left(5+10\RD+3\RD^2
\right)R^3\bfrac{\delta\sigma_*}{\sigma_*}^3, \qquad \textrm{for low $\treh$},\
\label{zetalowTR}
}
 where we assume that the perturbation of radiation,
 which has the origin from the inflaton field $\zeta_r$, is
 Gaussian and these higher order contributions are negligible, 
 compared to those of the curvaton.
Here $\RD$ is evaluated when the curvaton decay as
\dis{
\left. \RD\simeq \frac{3\rho_\sigma}{ 4\rho_r+3\rho_\sigma}
 \right|_{H=\Gamma_{\sigma}}
=\left.\frac{3r}{4+3r}\right|_{H=\Gamma_{\sigma}},
}
where $\rho_r$ and $\rho_\sigma$ are the energy densities of radiation and the 
curvaton respectively and $r$ is the ratio of them, $r\equiv\rho_\sigma/\rho_r $.
Note that $\zeta_r$ can be comparable to $R\zeta_\sigma$ with small $R$,
which becomes the general mixed inflaton-curvaton scenario~\cite{MixedIC}.

The nonlinearity parameters are given for a large initial amplitude, $\sigma_*\gg
\sv$, for high $\treh$ from \eq{zetahighTR} as
\dis{
  \fnl &= \bfrac{\tr}{1+\tr}^2\frac56\left(\frac{1}{\RD}-2-\RD \right)>-2\\
\gnl &= \bfrac{\tr}{1+\tr}^3\frac{25}{54}\left(-\frac{1}{\RD^2} -\frac{6}{\RD}
+2 +10\RD+3\RD^2 \right), \qquad\textrm{for high
  $\treh$},\label{nonGhighTR}
}
 and for a low $\treh$ from \eq{zetalowTR} as
\dis{
  \fnl &= \bfrac{\tr}{1+\tr}^2\frac56\left(-2-\RD \right)\\
\gnl &= \bfrac{\tr}{1+\tr}^3\frac{25}{54}\left(5 +10\RD+3\RD^2 \right), \qquad\textrm{for low
  $\treh$},\label{nonGlowTR}
}
The sizable large $\fnl \sim 100$ is obtained with a small ratio $\RD\sim
10^{-2}$ for a high $\treh$. However it is impossible to have 
 such a large non-Gaussianity for a low $\treh$ 
 since there are cancellations in the coefficients of inverse of $R$-terms. 
In this region, the curvaton starts the quartic oscillation,
 when the Universe is dominated by the oscillating inflaton field.

\begin{figure}[!t]
 \begin{tabular}{c c}
\includegraphics[width=0.5\textwidth,angle=0]{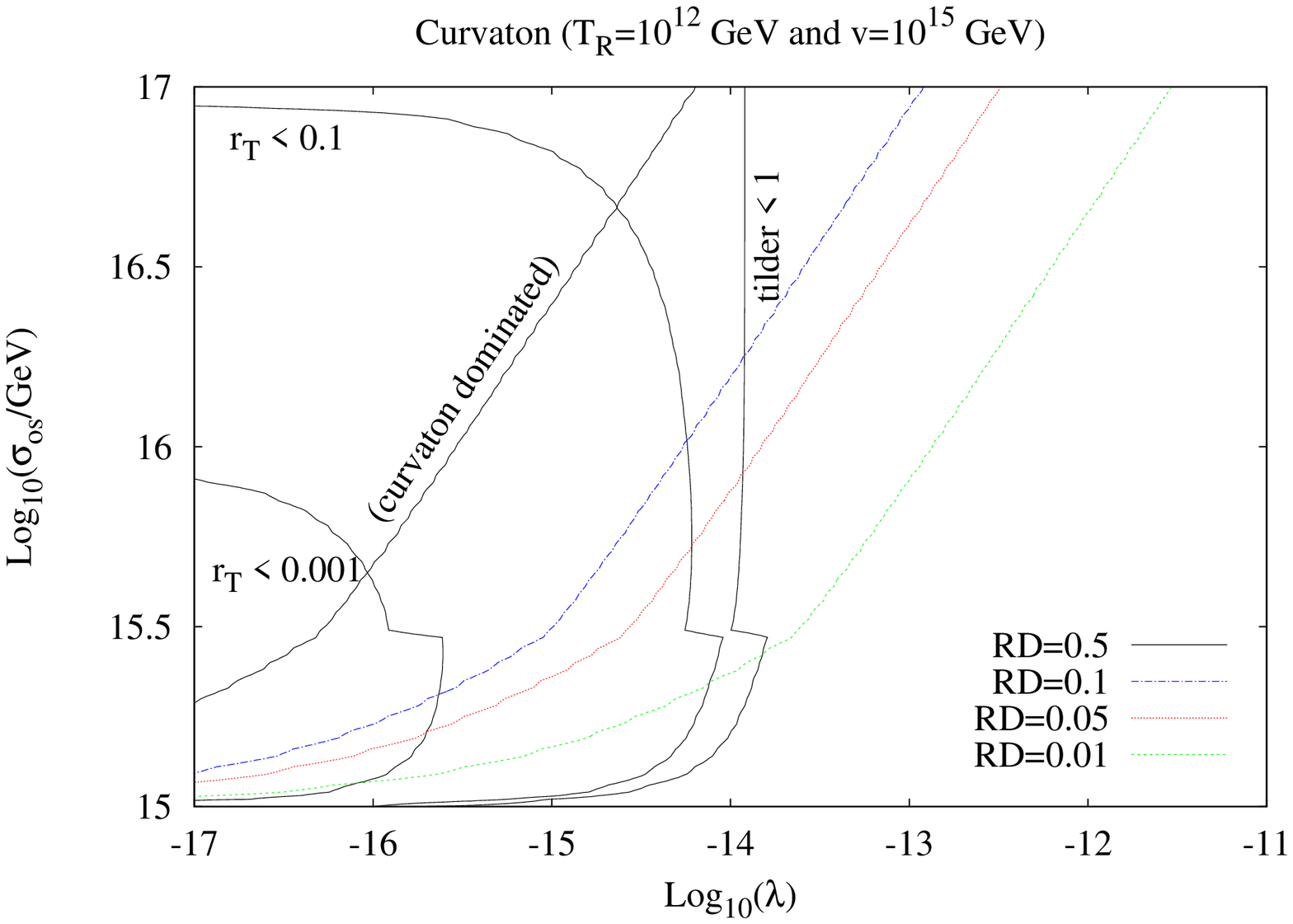} 
&
\includegraphics[width=0.5\textwidth,angle=0]{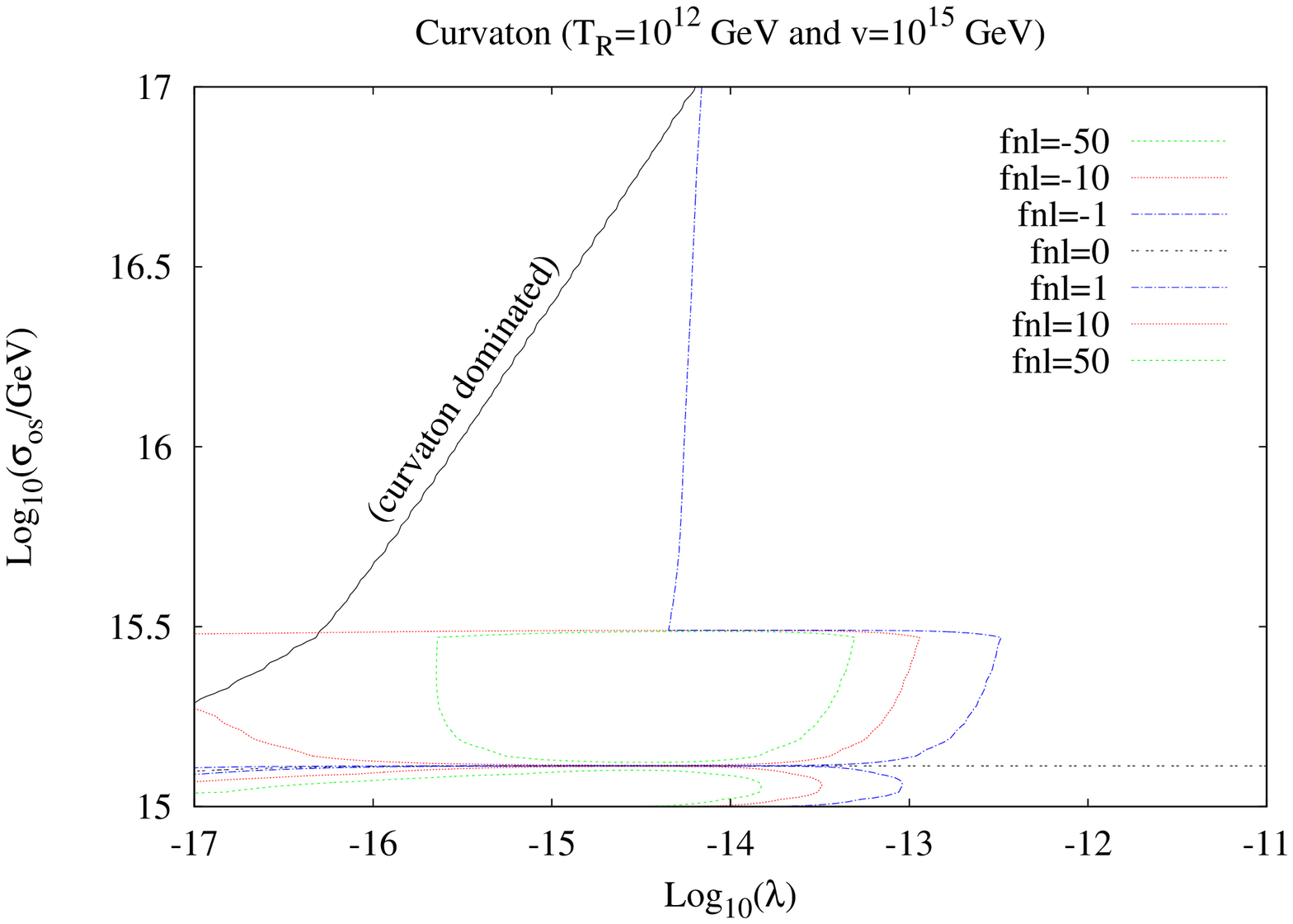} 
\end{tabular}
 \caption{Left: The contour plots of $\RD$ and $r_T$. 
Here we used $v=10^{15} \gev$ and $\treh=10^{12}\gev$.
Right: The contour plot of $\fnl$ in the same plane of the left.
For the sign of $\fnl$ refer to figure~\ref{fig:fnl2}.
} \label{fig:fnlTR12}
\end{figure}

\begin{figure}[!t]
 \begin{tabular}{c c}
\includegraphics[width=0.5\textwidth,angle=0]{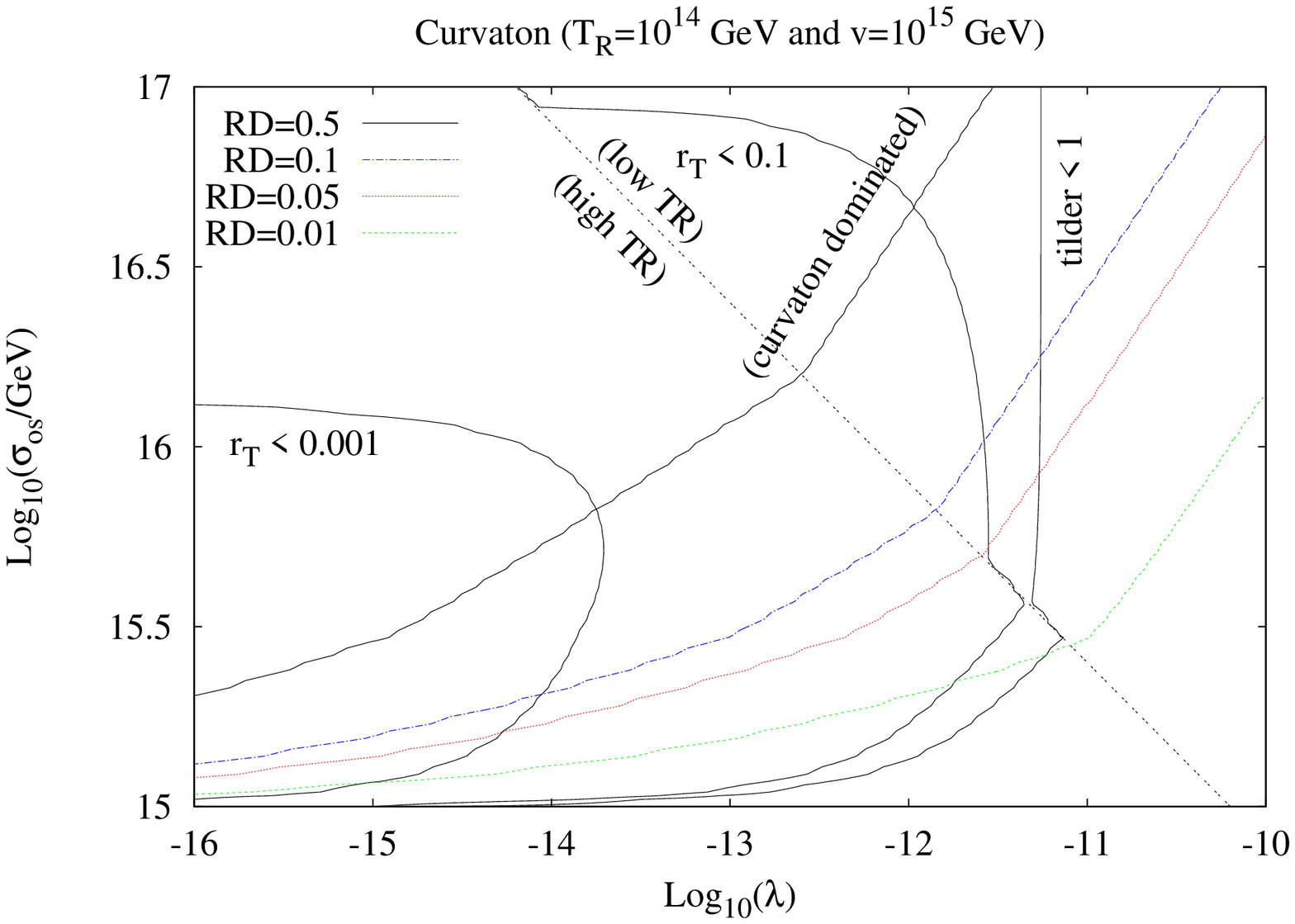} 
&
\includegraphics[width=0.5\textwidth,angle=0]{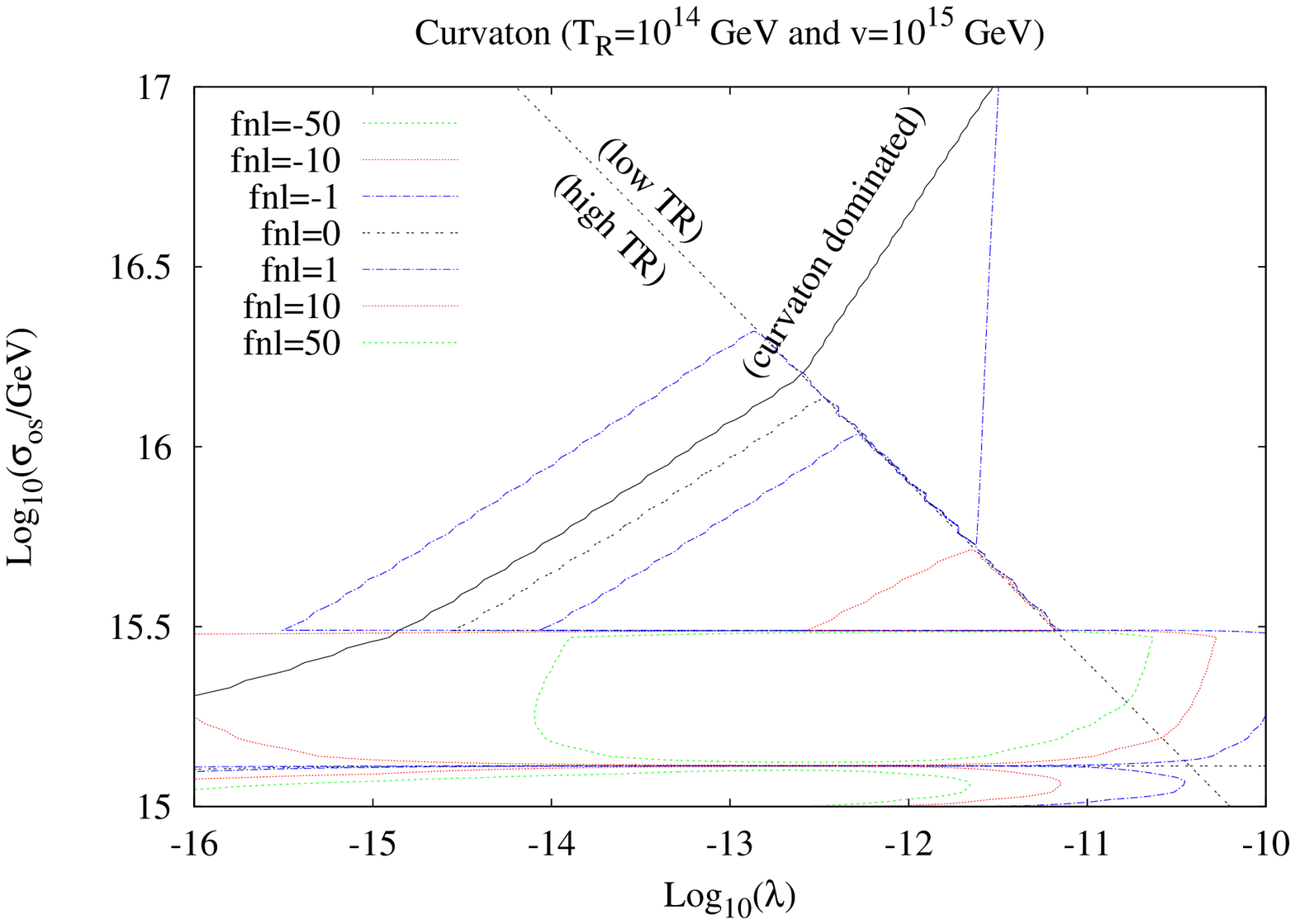} 
\end{tabular}
 \caption{Left: The contour plots of $\RD$ and $r_T$. 
Here we used $v=10^{15} \gev$ and $\treh=10^{14}\gev$.
Right: The contour plot of $\fnl$ in the same plane of the left.
For the sign of $\fnl$ refer to figure~\ref{fig:fnl2}.
} 
\label{fig:fnlTR14}
\end{figure}
\begin{figure}[!t]
 \begin{tabular}{c c}
\includegraphics[width=0.5\textwidth,angle=0]{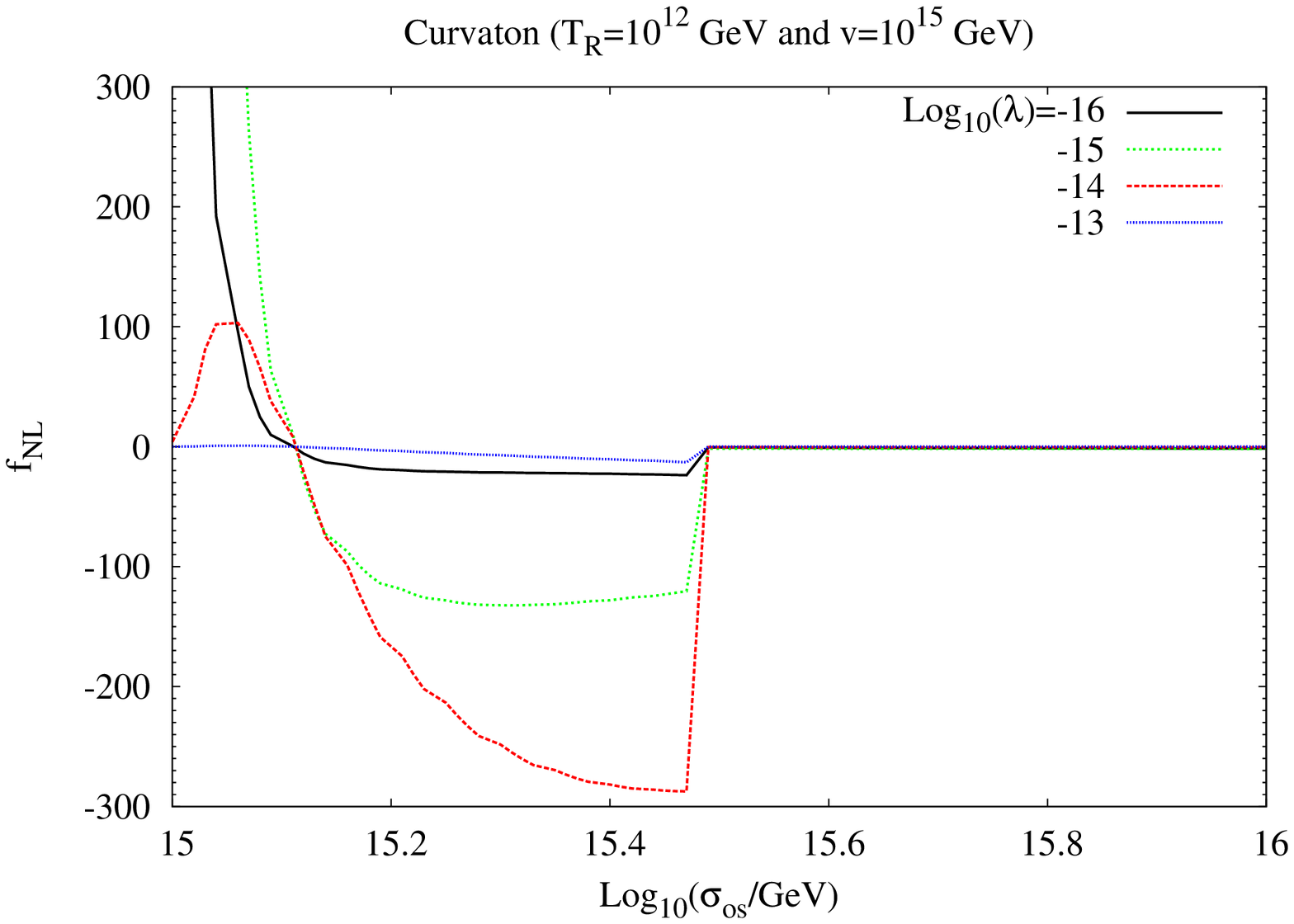} 
&
\includegraphics[width=0.5\textwidth,angle=0]{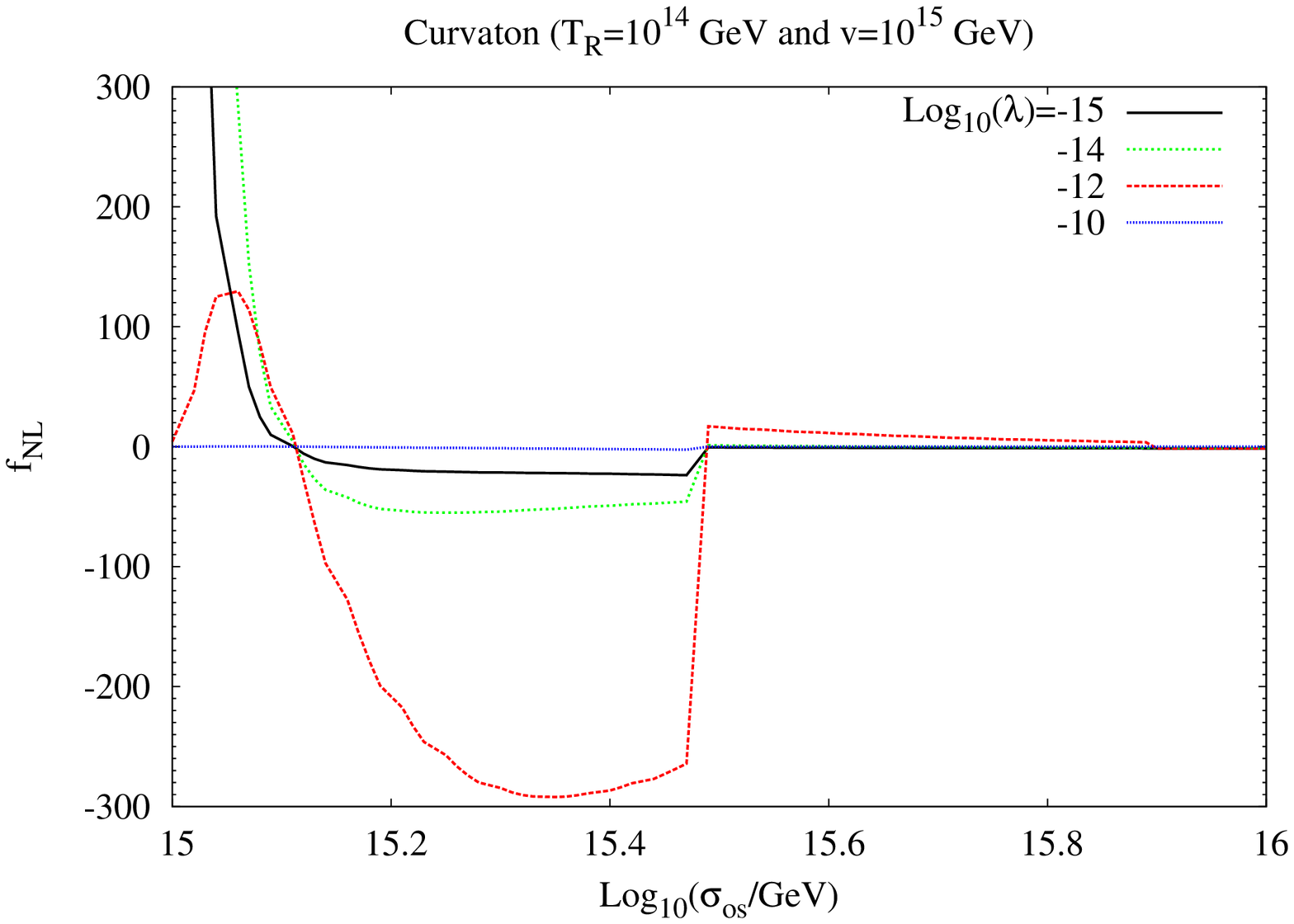} 
\end{tabular}
 \caption{The plots for $\fnl$ (left) with the fixed $\lambda$. 
Here $v=10^{15} \gev$ and $\treh=10^{12}\gev$ (left)
or $\treh=10^{14}\gev$ (right). }
 \label{fig:fnl2}
\end{figure}

In Figs~\ref{fig:fnlTR12} and~\ref{fig:fnlTR14},
 the contour plots of gravitational wave background and $\RD$ (left window)
 $\fnl$ (right window) are shown with observational constraints
 for the decay rate given  by Eq.~(\ref{Gamma:SaxionType})
 in the plane parameters of $\lambda$ and $\sigma_*$.
Here we have fixed VEVs $v=10^{15} \gev$, and the reheating temperature
 $\treh=10^{12}\gev$ (figure~\ref{fig:fnlTR12}) 
 and $\treh=10^{14}\gev$ (figure~\ref{fig:fnlTR14}) separately.
The observational constraints include the tensor-to-scalar ratio 
$r_T = 0.1 $ and $10^{-3}$ for the expected sensitivity of B-mode detection
 by Planck and future instruments such as CMBPol~\cite{CMBPol}, respectively.
For given $\sigma_{\rm os}$ and $\lambda$, imposing \eq{Pzetawmap} on \eq{Pzeta}
 determines $H_*$ with $\epsilon\simeq0.02$ from \eq{ns}.

As can be seen in the figures, 
 the large $\sigma_*$  or the large $\lambda$ region is excluded 
 by the null detection of gravitational wave.
The region with $\tilde{r} <1$ belongs to this excluded region,
which means that in the allowed region the power spectrum is dominated by
that from the curvaton in our scenario.
Above the diagonal line, written above (curvaton dominated), the curvaton energy
dominates the Universe before it decays. 
In the limit of curvaton domination ($\RD=1$)  the 
non-linearity parameter becomes $\fnl=-5/3$ or $-5/2$ for high $\treh$
and low $\treh$ respectively.
Below the line '(curvaton dominated)', $\RD$ can be much smaller than $1$ and
 thus there is a chance to obtain large non-Gaussianity
for high $\treh$ case.
This region appears in figure~\ref{fig:fnlTR14} (Right window),
as a wedge shape above $\sos>\sv$.
In this region, it is possible 
 to generate large non-Gaussianity of the order of 20.
On the other hand, for a larger initial amplitude the curvaton oscillation 
starts before reheating, corresponding to low-$\treh$, and the non-Gaussianity
is significantly suppressed because of the cancellation as discussed after Eq.~(\ref{nonGlowTR}).
This happens in the '(low $\treh$)' region above the dashed diagonal
line in figure~\ref{fig:fnlTR14}.
For $\treh=10^{12}\gev$ (figure~\ref{fig:fnlTR12}), all the drawn region
corresponds to '(low $\treh$)' and the non-Gaussianity is small
for $\sigma_{\rm os} > \sigma_v$.

The interesting behavior happens for the initial amplitude $v<\sstar<\sv$,
where the oscillation of the curvaton starts in the quadratic term dominated potential.
There are two regions depending on the sign of $\fnl$,
positive for $v < \sstar\lesssim 1.8v$ and negative for $1.8v < \sstar\lesssim
\sv$. This difference is due to the evolving effective mass in a double well
potential, which is constant in the pure quadratic potential.
In the figure~\ref{fig:fnl2} we show the plot of $\fnl$ depending on $\sstar$ 
for given $\lambda$ for $\treh=10^{12}\gev$ (left) and $10^{14}\gev$ (right)
respectively.
The positive $\fnl$ with the magnitude of the order of 100
is possible for a small initial amplitude $v<\sstar\lesssim 1.8 v$.

\section{Summary}
\label{summary}

We have studied the density perturbation generated by a curvaton 
 whose potential is flat with small self-coupling in a double well type.
We have used a large VEV, $v > 10^{15}$ GeV, to avoid the domain wall formation 
and a larger initial curvaton amplitude $\sigma_*$ which is
easily obtained in this flat potential.

We have analyzed the cosmological evolution of the scalar field in this flat double
 well potential to see the viability of the field as a curvaton to generate 
 the primordial density perturbation to explain the structure formation
 and the anisotropies in the CMB.
With a large initial expectation value $\sigma_{\rm os} \gg v$,
 the energy density decreases as initially $\propto a^{-4}$ and
 at late time as $\propto a^{-3}$.
This is same as a massive curvaton with self-interaction with vanishing VEV.
However, three crucial differences and features appear.
One is the tachyonic amplification of the fluctuation 
 by the negative mass squared at the origin,  
 for a particular initial value $\sigma_{\rm os}$
 which periodically appear in the parameter space.
We have shown that, except for these tuned boundaries, 
 the density perturbation and other nonlinear parameters
 are well approximated by analytic formula.
The second is the suppression of  non-Gaussianity
 even for very subdominant curvaton,
 if the reheating temperature after inflation is as low as
 not to satisfy Eq.~(\ref{highTR}).
The other is the non-trivial behavior of $\fnl$ even when the curvaton
field starts oscillation trapped at one of the minima.
We found a successful scenario needs the flat potential with small self-coupling of the
 field of the order of $\lesssim 10^{-10}$ for a reasonable reheating temperature. 

In conclusion, 
 we have found the differences in massive self-interacting curvaton models
 with and without VEV.
In addition, in a double well potential curvaton model,
 non-linear parameters can not be so large for a large initial field value
 and a low reheating temperature.
Therefore, if both large nonlinearity and B-mode polarization will be detected,
 the potential of the curvaton or thermal history of the early Universe
 will be constrained.

%
\section*{Acknowledgments}
We would like to thank Tomo Takahashi for valuable communication.
This work is in part supported by 
 the Korea Research Foundation Grant funded
 from the Korean Government (KRF-2008-341-C00008),
 by the second stage of Brain Korea 21 Project in 2006 (K.Y.C), 
 and by the scientific research grants from Hokkai-Gakuen (O.S).
O.S would like to thank Department of Physics at Pusan National University 
 for their warm hospitality where many parts of this work have been done.




\begin{thebibliography}{99}

\bibitem{Inflation}
  A.~A.~Starobinsky,
  JETP Lett.\  {\bf 30} 682 (1979)
  [Pisma Zh.\ Eksp.\ Teor.\ Fiz.\  {\bf 30} 719 (1979)];\\
  K.~Sato,
  Mon.\ Not.\ Roy.\ Astron.\ Soc.\  {\bf 195}, 467 (1981);\\
  A.~H.~Guth,
  Phys.\ Rev.\  D {\bf 23}, 347 (1981); \\
  A.~D.~Linde,
  Phys.\ Lett.\  B {\bf 108} 389 (1982);\\
  A.~Albrecht and P.~J.~Steinhardt,
  Phys.\ Rev.\ Lett.\  {\bf 48} 1220 (1982).




\bibitem{InflationFluctuation}
  S.~W.~Hawking,
  Phys.\ Lett.\  B {\bf 115}, 295 (1982); \\
  A.~A.~Starobinsky,
  Phys.\ Lett.\  B {\bf 117}, 175 (1982); \\
  A.~H.~Guth and S.~Y.~Pi,
  Phys.\ Rev.\ Lett.\  {\bf 49}, 1110 (1982).

\bibitem{Maldacena:2002vr}
  J.~M.~Maldacena,
  JHEP {\bf 0305}, 013 (2003).

\bibitem{Komatsu:2010fb}
  E.~Komatsu {\it et al.},
  arXiv:1001.4538 [astro-ph.CO].

\bibitem{Planck} 
  http://www.rssd.esa.int/index.php?project=PLANCK

\bibitem{Chen:2006xjb}
  X.~Chen, R.~Easther and E.~A.~Lim,
  JCAP {\bf 0706}, 023 (2007).

\bibitem{DBI}
  E.~Silverstein and D.~Tong,
  Phys.\ Rev.\  D {\bf 70}, 103505 (2004); \\
  M.~Alishahiha, E.~Silverstein and D.~Tong,
  Phys.\ Rev.\  D {\bf 70}, 123505 (2004); \\
  X.~Chen, M.~X.~Huang, S.~Kachru and G.~Shiu,
  JCAP {\bf 0701}, 002 (2007).


\bibitem{Choi:2007su}
  K.~Y.~Choi, L.~M.~H.~Hall and C.~van de Bruck,
  JCAP {\bf 0702} 029 (2007).

\bibitem{Zaldarriaga:2003my}
  M.~Zaldarriaga,
  Phys.\ Rev.\  D {\bf 69}, 043508 (2004).

\bibitem{Byrnes:2008wi}
  C.~T.~Byrnes, K.~Y.~Choi and L.~M.~H.~Hall,
  JCAP {\bf 0810} 008 (2008).

\bibitem{Byrnes:2008zy}
  C.~T.~Byrnes, K.~Y.~Choi and L.~M.~H.~Hall,
  JCAP {\bf 0902} 017 (2009).

\bibitem{Lyth:2005qk}
  D.~H.~Lyth,
  JCAP {\bf 0511} 006 (2005).

\bibitem{Alabidi:2006wa}
  L.~Alabidi and D.~Lyth,
  JCAP {\bf 0608} 006 (2006).



\bibitem{Sasaki:2008uc}
  M.~Sasaki,
  Prog.\ Theor.\ Phys.\  {\bf 120} 159 (2008).

\bibitem{Naruko:2008sq}
  A.~Naruko and M.~Sasaki,
  Prog.\ Theor.\ Phys.\  {\bf 121}, 193 (2009).

\bibitem{preheating}
  K.~Enqvist, A.~Jokinen, A.~Mazumdar, T.~Multamaki and A.~Vaihkonen,
  Phys.\ Rev.\ Lett.\  {\bf 94} 161301 (2005); JCAP {\bf 0503} 010 (2005); \\
  A.~Jokinen and A.~Mazumdar,
  JCAP {\bf 0604} 003 (2006).



\bibitem{Byrnes:2010em}
  for a recent review on local type non-Gaussianity, see e.g., 
  C.~T.~Byrnes and K.~Y.~Choi,
  arXiv:1002.3110 [astro-ph.CO];
  D.~Wands,
  Class.\ Quant.\ Grav.\  {\bf 27}, 124002 (2010).
  

\bibitem{Mollerach:1989hu}
  S.~Mollerach,
  Phys.\ Rev.\  D {\bf 42} 313 (1990).

\bibitem{Linde:1996gt}
  A.~D.~Linde and V.~F.~Mukhanov,
  Phys.\ Rev.\  D {\bf 56} 535 (1997).

\bibitem{CurvatonLW}
  D.~H.~Lyth and D.~Wands,
  Phys.\ Lett.\  B {\bf 524}, 5 (2002).

\bibitem{CurvatonMT}
  T.~Moroi and T.~Takahashi,
  Phys.\ Lett.\  B {\bf 522}, 215 (2001)
  [Erratum-ibid.\  B {\bf 539}, 303 (2002)].

\bibitem{CurvatonES}
  K.~Enqvist and M.~S.~Sloth,
  Nucl.\ Phys.\  B {\bf 626}, 395 (2002).

\bibitem{CurvatonNG}
  D.~H.~Lyth, C.~Ungarelli and D.~Wands,
  Phys.\ Rev.\  D {\bf 67}, 023503 (2003).




\bibitem{GWbackground}
  B.~Allen,
  Phys.\ Rev.\  D {\bf 37}, 2078 (1988); \\
  V.~Sahni,
  Phys.\ Rev.\  D {\bf 42}, 453 (1990).

\bibitem{Nakayama:2009ce}
  K.~Nakayama and J.~Yokoyama,
  JCAP {\bf 1001}, 010 (2010).

\bibitem{DECIGO} 
  N.~Seto, S.~Kawamura and T.~Nakamura,
  Phys.\ Rev.\ Lett.\  {\bf 87}, 221103 (2001).

\bibitem{BBO}
  J.~Crowder and N.~J.~Cornish,
  Phys.\ Rev.\  D {\bf 72}, 083005 (2005).




\bibitem{Bartolo:2003jx}
  N.~Bartolo, S.~Matarrese and A.~Riotto,
  Phys.\ Rev.\  D {\bf 69} 043503 (2004).

\bibitem{Lyth:2005fi}
  D.~H.~Lyth and Y.~Rodriguez,
  Phys.\ Rev.\ Lett.\  {\bf 95} 121302 (2005).


\bibitem{Sasaki:2006kq}
  M.~Sasaki, J.~Valiviita and D.~Wands,
  Phys.\ Rev.\  D {\bf 74} 103003 (2006).

\bibitem{Malik:2006pm}
  K.~A.~Malik and D.~H.~Lyth,
  JCAP {\bf 0609} 008 (2006).


\bibitem{MixedIC}
  D.~Langlois and F.~Vernizzi,
  Phys.\ Rev.\  D {\bf 70}, 063522 (2004); \\
  G.~Lazarides, R.~R.~de Austri and R.~Trotta,
  Phys.\ Rev.\  D {\bf 70}, 123527 (2004); \\
  F.~Ferrer, S.~Rasanen and J.~Valiviita,
  JCAP {\bf 0410}, 010 (2004); \\
  T.~Moroi, T.~Takahashi and Y.~Toyoda,
  Phys.\ Rev.\  D {\bf 72} 023502 (2005);\\
  T.~Moroi and T.~Takahashi,
  Phys.\ Rev.\  D {\bf 72} 023505 (2005); \\
  K.~Ichikawa, T.~Suyama, T.~Takahashi and M.~Yamaguchi,
  Phys.\ Rev.\  D {\bf 78}, 023513 (2008).




\bibitem{Dimopoulos:2003ss}
  K.~Dimopoulos, G.~Lazarides, D.~Lyth and R.~Ruiz de Austri,
  Phys.\ Rev.\  D {\bf 68} 123515 (2003).

\bibitem{Huang:2008zj}
  Q.~G.~Huang,
  JCAP {\bf 0811}, 005 (2008).

\bibitem{Chingangbam:2009xi}
  P.~Chingangbam and Q.~G.~Huang,
  JCAP {\bf 0904}, 031 (2009).

\bibitem{Enqvist:2009zf}
  K.~Enqvist, S.~Nurmi, G.~Rigopoulos, O.~Taanila and T.~Takahashi,
  JCAP {\bf 0911}, 003 (2009).


  
\bibitem{Choi:2007fya}
  K.~Y.~Choi and J.~O.~Gong,
  JCAP {\bf 0706} 007 (2007).

\bibitem{Assadullahi:2007uw}
  H.~Assadullahi, J.~Valiviita and D.~Wands,
  Phys.\ Rev.\  D {\bf 76} 103003 (2007).

\bibitem{Huang:2008rj}
  Q.~G.~Huang,
  JCAP {\bf 0809} 017 (2008).



\bibitem{Enqvist:2005pg}
  K.~Enqvist and S.~Nurmi,
  JCAP {\bf 0510} 013 (2005).

\bibitem{Enqvist:2008gk}
  K.~Enqvist and T.~Takahashi,
  JCAP {\bf 0809}, 012 (2008).

\bibitem{Huang:2008bg}
  Q.~G.~Huang and Y.~Wang,
  JCAP {\bf 0809} 025 (2008).

\bibitem{Enqvist:2009eq}
  K.~Enqvist and T.~Takahashi,
  JCAP {\bf 0912}, 001 (2009).

\bibitem{Enqvist:2009ww}
  K.~Enqvist, S.~Nurmi, O.~Taanila and T.~Takahashi,
  JCAP {\bf 1004}, 009 (2010).

\bibitem{Byrnes:2010xd}
  C.~T.~Byrnes, K.~Enqvist and T.~Takahashi,
  arXiv:1007.5148 [astro-ph.CO].



\bibitem{Linde:1991km}
  A.~D.~Linde,
  Phys.\ Lett.\  B {\bf 259} 38 (1991).

\bibitem{Kasuya:1996ns}
  S.~Kasuya, M.~Kawasaki and T.~Yanagida,
  Phys.\ Lett.\  B {\bf 409}, 94 (1997); \\
  S.~Kasuya and M.~Kawasaki,
  Phys.\ Rev.\  D {\bf 56}, 7597 (1997); 
  Phys.\ Rev.\  D {\bf 58}, 083516 (1998).


\bibitem{Kawasaki:2008mc}
  M.~Kawasaki, K.~Nakayama and F.~Takahashi,
  JCAP {\bf 0901} 026 (2009).



\bibitem{KSVZ}
  J.~E.~Kim,
  Phys.\ Rev.\ Lett.\  {\bf 43}, 103 (1979);\\
  M.~A.~Shifman, A.~I.~Vainshtein and V.~I.~Zakharov,
  Nucl.\ Phys.\  B {\bf 166}, 493 (1980).

\bibitem{Dimopoulos:2003ii}
  K.~Dimopoulos, G.~Lazarides, D.~Lyth and R.~Ruiz de Austri,
  JHEP {\bf 0305}, 057 (2003).



\bibitem{Kofman:1995fi}
  L.~Kofman, A.~D.~Linde and A.~A.~Starobinsky,
  Phys.\ Rev.\ Lett.\  {\bf 76} 1011 (1996).


\bibitem{Lyth:2004gb}
  D.~H.~Lyth, K.~A.~Malik and M.~Sasaki,
  JCAP {\bf 0505} 004 (2005).
  
\bibitem{Langlois:2008vk}
  D.~Langlois, F.~Vernizzi and D.~Wands,
  JCAP {\bf 0812} 004 (2008).




\bibitem{CMBPol}
  D.~Baumann {\it et al.}  [CMBPol Study Team Collaboration],
  AIP Conf.\ Proc.\  {\bf 1141}, 10 (2009).








  












\end{thebibliography}
\end{document}